\documentclass[showpacs, oneside, twocolumn, prd, amsmath, amssymb, nofootinbib, superscriptaddress]{revtex4-1}

\pdfoutput=1
\usepackage{cases}
\usepackage{amsmath}
\usepackage{amssymb}
\usepackage{amsfonts}
\usepackage{amssymb}
\usepackage{dcolumn}
\usepackage{bm}
\usepackage{bbm}
\usepackage{graphicx}
\usepackage{xcolor}
\usepackage{array}
\usepackage{subfigure}
\usepackage{hyperref}
\usepackage{wasysym}
\usepackage{mathrsfs}

\newcommand{\be}{\begin{equation}}
\newcommand{\ee}{\end{equation}}
\newcommand{\ba}{\begin{eqnarray}}
\newcommand{\ea}{\end{eqnarray}}

\newcommand{\gsim}{\mathrel{\hbox{\rlap{\lower.55ex \hbox {$\sim$}}
			\kern-.3em \raise.4ex \hbox{$>$}}}}
\newcommand{\lsim}{\mathrel{\hbox{\rlap{\lower.55ex \hbox {$\sim$}}
			\kern-.3em \raise.4ex \hbox{$<$}}}}

\hypersetup{colorlinks=true,
	breaklinks=true,
	pdfstartview=Fit,
	linkcolor=blue,
	citecolor=blue,
	urlcolor=blue}

\begin{document}

\title{A Compaction Function Analysis of CMB $\mu$ distortion Constraints on Primordial Black Holes}

\author{Junyue Yang}
%\email{sa24022015@mail.ustc.edu.cn}
\affiliation{Department of Astronomy, School of Physical Sciences, University of Science and Technology of China, Hefei, Anhui 230026, China}
\affiliation{CAS Key Laboratory for Researches in Galaxies and Cosmology/Department of Astronomy, School of Astronomy and Space Science, University of Science and Technology of China, Hefei, Anhui 230026, China}
\affiliation{School of Physics Science And Technology, Wuhan University, No.299 Bayi Road, Wuhan, Hubei, China}
\affiliation{School of Physics, The University of Electronic Science and Technology of China, 88 Tian-run Road, Chengdu, China}

\author{Xiaoding Wang}
%\email{ylxdxx@qq.com}
\affiliation{The Key Laboratory of Cosmic Rays (Tibet University), Ministry of Education, Lhasa 850000, Tibet, China}

\author{Xiao-Han Ma}
%\email{mxh171554@mail.ustc.edu.cn}
\affiliation{Department of Astronomy, School of Physical Sciences, University of Science and Technology of China, Hefei, Anhui 230026, China}
\affiliation{CAS Key Laboratory for Researches in Galaxies and Cosmology/Department of Astronomy, School of Astronomy and Space Science, University of Science and Technology of China, Hefei, Anhui 230026, China}
\affiliation{Kavli Institute for the Physics and Mathematics of the Universe (WPI), UTIAS, The University of Tokyo, Chiba 277-8583, Japan}

\author{Dongdong Zhang}
%\email{don@mail.ustc.edu.cn}
\affiliation{Department of Astronomy, School of Physical Sciences, University of Science and Technology of China, Hefei, Anhui 230026, China}
\affiliation{CAS Key Laboratory for Researches in Galaxies and Cosmology/Department of Astronomy, School of Astronomy and Space Science, University of Science and Technology of China, Hefei, Anhui 230026, China}
\affiliation{Kavli Institute for the Physics and Mathematics of the Universe (WPI), UTIAS, The University of Tokyo, Chiba 277-8583, Japan}

\author{\\Sheng-Feng Yan}
\email{sfyan@uestc.edu.cn}
\affiliation{School of Physics, The University of Electronic Science and Technology of China, 88 Tian-run Road, Chengdu, China}

\author{Amara Ilyas}
\email{aarks@ustc.edu.cn}
\affiliation{Department of Astronomy, School of Physical Sciences, University of Science and Technology of China, Hefei, Anhui 230026, China}
\affiliation{CAS Key Laboratory for Researches in Galaxies and Cosmology/Department of Astronomy, School of Astronomy and Space Science, University of Science and Technology of China, Hefei, Anhui 230026, China}

\author{Yi-Fu Cai}
\email{yifucai@ustc.edu.cn}
\affiliation{Department of Astronomy, School of Physical Sciences, University of Science and Technology of China, Hefei, Anhui 230026, China}
\affiliation{CAS Key Laboratory for Researches in Galaxies and Cosmology/Department of Astronomy, School of Astronomy and Space Science, University of Science and Technology of China, Hefei, Anhui 230026, China}

\begin{abstract}
Primordial black holes (PBHs) are considered viable candidates for dark matter and the seeds of supermassive black holes (SMBHs), with their fruitful physical influences providing significant insights into the conditions of the early Universe. Cosmic microwave background (CMB) $\mu$ distortion tightly constrain the abundance of PBHs in the mass range of $10^4 \sim 10^{11} M_{\odot}$ recently, limiting their potential to serve as seeds for the SMBHs observed. Given that $\mu$ distortion directly constrain the primordial power spectrum, it is crucial to employ more precise methods in computing PBH abundance to strengthen the reliability of these constraints. By a Press-Schechter (PS) type method utilizing the compaction function, we find that the abundance of PBHs could be higher than previously estimated constraints from $\mu$ distortion observations. Furthermore, our analysis shows that variations in the shape of the power spectrum have a negligible impact on our conclusions within the mass ranges under consideration. This conclusion provides us a perspective for further research on the constrain of PBH by $\mu$ distortion.
\end{abstract}

%\pacs{04.50.Kd, 98.80.-k, 95.36.+x, 98.80.Es}

% 04.50.Kd Modified theories of gravity
% 98.80.?k Cosmology
% 95.36.+x  Dark energy
% 98.80.Es Observational cosmology

\maketitle

\section{Introduction}

It has been proposed that PBHs can be generally produced from the collapse of primordial cosmological perturbations in the very early Universe  \cite{Zeldovich:1967lct, hawking_gravitationally_1971,carr_black_1974,carr1975primordial,khlopov_primordial_2010}. They have become a central topic of interest in cosmological research because PBHs could play a role in various scenarios, such as constituting dark matter or the source of gravitational wave signals \cite{Chapline:1975ojl,MACHO:1996qam,Bird:2016dcv,Hall:2020daa,DeLuca:2020fpg,Franciolini:2021tla,LIGOScientific:2021job,NANOGrav:2023hvm,Chen:2021nxo,Cai:2021yvq,Cai:2022kbp,Cai:2023dls}.
Moreover, the abundance and mass function of PBHs are related to the statistical properties of the primordial perturbations \cite{Byrnes:2018txb,Pi:2022ysn,Harada:2013epa,Carr:2023tpt,Escriva:2022duf}. When we assume that the primordial perturbations follow a Gaussian distribution, the power spectrum of primordial perturbations determines the abundance and mass function of PBHs. Consequently, the PBH abundance and mass function are connected with cosmic origin models, making PBHs a powerful probe for the early Universe \cite{Ivanov:1994pa,Garcia-Bellido:1996mdl,Quintin:2016qro,Hertzberg:2017dkh,Ballesteros:2017fsr,Franciolini:2018vbk,Pi:2017gih,Chen:2020uhe,carr_constraints_2021,Cai:2018tuh,Cai:2020ovp,Cai:2021wzd,Cai:2023ptf}. Besides these motivations, PBHs may also be the seeds of SMBHs in our Universe \cite{kawasaki_primordial_2012}. 
Unlike black holes formed from stellar collapse, PBHs exhibit a wide mass distribution, so the SMBHs seeding from PBHs can gain most of their mass originally instead of acquiring long-term accretion \cite{Hooper_2024}. As many SMBHs at high redshift ($z \simeq 7 \sim 10$) with the mass of $10^7 \sim 10^{10}M_{\odot}$ observed by the James Webb Space Telescope (JWST) and Atacama Large Millimeter/submillimeter Array (ALMA) recently, PBH can be a possible solution for the mystery of their birth \cite{labbe_population_2023,10.1093/mnras/stad266,Liu:2023ymk}. 
However, the possible abundance of PBHs has been constrained by various direct and indirect observations over decades of study. Among these, the constraints from CMB distortion, an indirect constraint, are significant for the mass range we are concerned with \cite{Carr:1993aq}.

CMB distortions are deviations of the photon distribution from an ideal blackbody radiation spectrum, providing insights into physical processes during the early Universe \cite{khatri_creation_2012,khatri_beyond_2012}. 
Energy inputs shift the photon distribution from a blackbody spectrum to a Bose-Einstein distribution with a chemical potential, resulting in $\mu$-type distortion \cite{sunyaev_small_1970,chluba_probing_2012}.
Various processes lead to the energy injection into the isotropic part of the radiation field. One is damping small-scale perturbations, which liberate energy stored as sound waves \cite{Lifshitz:1945du,Lifshitz:1963ps,silk_cosmic_1968}.
The observational limit of $\mu$-type distortion has been established at $9\times10^{-5}$ by the COBE Far Infrared Absolute Spectrophotometer (FIRAS) instrument, providing constraints for the primordial power spectrum and cosmological models \cite{fixsen_cosmic_1996,fixsen_spectrum_1998}. 
Further, the constraints of the power spectrum of primordial curvature perturbations by $\mu$ distortion can be extended to constraints on the abundance of PBHs within the mass range approximately $10^4\sim10^{11} M_{\odot}$ as the candidate for SMBH seeds \cite{kohri_testing_2014, PhysRevD.94.103522,PhysRevD.100.103521,abitbol_prospects_2017, García-Bellido_2017,Deng_2021,Acharya_2020,yang_constraints_2022,PhysRevD.105.103535}. 

In this paper, we delve deeper into the calculation methodologies associated with PBH abundance to improve the results comprehensively. It has been noted the curvature perturbation, which is traditionally used in driving the constraints from distortions, may not be the ideal parameter for assessing PBH formation \cite{de_luca_note_2022}. The compaction function, defined as the smoothed density contrast within a region, has recently been commonly used to estimate the abundance of PBHs more accurately than the curvature perturbation \cite{PhysRevD.101.044022,Escrivà_2021,PhysRevD.103.063538,PhysRevD.91.084057,PhysRevLett.122.141302, Young_2019}. On the other hand, the threshold for PBH formation can be significantly influenced by the shape of the perturbation profile according to numerical simulations \cite{musco_threshold_2019}.  In this work, we use a Press-Schechter-type method based on the compaction function to connect the primordial power spectrum and the abundance of PBHs, thereby updating the constraints imposed by $\mu$ distortion on PBHs.

This paper is organized as follows: Section.~\ref{sec:section2} begins with a detailed review of the methodologies used to constrain the abundance of PBHs through the upper limits provided by CMB distortions. This includes the approaches for calculating PBH abundance based on curvature perturbation, examining the mechanisms generating CMB distortions, and the constraints imposed by the current observations on the $\mu$ distortion. Subsequently, we extensively analyze PBH abundance constraints utilizing the compaction function for primordial power spectrum with three types of feature: $\delta$, lognormal, and box form. 
In Section.~\ref{sec:section3}, we present the results derived from applying the compaction function, which relaxed the constraints on PBH abundance by at least 1-2 orders of magnitude. 
The difference between the various types of power spectrum leaves slight effects on the constraints on PBH abundance, and the difference is less than 1 order of magnitude. The paper concludes with Section.~\ref{sec:section4}, where we summarize our results and further discuss the implications to the physics of PBHs.

\section{Methodology}\label{sec:section2}

This section systematically reviews methods to constrain the abundance of PBHs using CMB distortion observations.
The analysis primarily utilizes curvature perturbations to estimate PBH abundance \cite{chluba_probing_2012,kohri_testing_2014,nakama_limits_2018}.
We also discuss a more precise method based on the compaction function, strengthening the reliability of the constraints on PBH abundance from CMB distortions.

\subsection{Abundance of PBHs based on curvature perturbation}

PS formalism, which estimates the fraction of overdense regions in a perturbed density field, was initially used to calculate the halo mass function for evaluating galaxy formation \cite{Press:1973iz}. It has since been widely applied to estimate the abundance of PBHs.
In this formalism, the abundance of PBHs is proportional to the integral of the probability density function (PDF) that exceeds a certain threshold \cite{green_new_2004,young_calculating_2014,kitajima_primordial_2021}. Current cosmological observations suggest that the curvature perturbations $\zeta$ follow a Gaussian distribution well on large scales \cite{sureda_press-schechter_2021,Planck:2018vyg}.

When the perturbations re-enter the Hubble radius, only a few overdense perturbations located in the tail regime of the PDF, where $\zeta > \zeta_c$ (the threshold is approximately 0.67), may collapse into black holes \cite{nakama_limits_2018}. Roughly we can establish a relation between the mass of a PBH and its formation time, specifically, the time of horizon re-entry and subsequent collapse, expressed as
\begin{equation}
M_{\rm{PBH}}=\gamma M_{\mathrm{H}}\bigg|_{\text {at formation }}=\frac{\gamma}{2G}H^{-1}_{\rm{form}}.
\end{equation}
It implies that the mass of PBHs is proportional to the Hubble scale at formation. 
$\gamma$ is a correction factor that denotes the fraction of an overdense region that collapses into a black hole. It depends on the background and is typically around 0.2 \cite{carr1975primordial}.

In the simplest case, the PDF of the density contrast $\delta$ is usually assumed to be Gaussian \cite{Carr:2009jm,sasaki_primordial_2018},
\begin{equation}
    P(\delta)=\frac{1}{\sqrt{2\pi \sigma^2}}\rm{exp}\left(\frac{-\delta^2}{2\sigma^2}\right),
\end{equation} 
where the variance of perturbations $\sigma$ can be expressed in terms of the corresponding scales $k$ of PBH mass as:
\begin{equation}
\begin{aligned}
    \sigma^2\left[M_{\rm{PBH}}(k)\right]&=\int \frac{\mathrm{d}q}{q} \mathcal{P}_{\delta}(q)W^2(q/k) \\&=\int \frac{\mathrm{d}q}{q}\frac{16}{81}\left(\frac{q}{k}\right)^4 \mathcal{P}_{\zeta}(q)W^2\left(\frac{q}{k}\right).\label{variance}
    \end{aligned}
\end{equation}
The window function is Gaussian $W(x)=\exp(-x^2/2)$ to represent the perturbation smoothing and the relation between density and contrast curvature perturbation $\delta\simeq \frac{4}{9}(\frac{k}{aH})^2 \zeta$ is applied.
$\mathcal{P}_{\delta}(k)$ and $\mathcal{P}_\zeta (k)$ represent the dimensionless power spectrum of density contrast and curvature perturbation, respectively. 
Therefore, the mass function of PBHs at the formation epoch can be given by
\begin{equation} \label{eq:PBHbeta}
    \beta(M_{\rm{PBH}})=\frac{1}{\sqrt{2\pi }\sigma(M_{\rm{PBH}})    }\int^{\infty}_{\delta_c}  \mathrm{d} \delta \mathrm{exp} \left[ \frac{-\delta^2}{2\sigma^2(M_{\rm{PBH}})} \right].
\end{equation}
It is related to the current fraction of PBHs against total dark matter $f_{\mathrm{PBH}}$ as
\begin{equation}\label{eq_beta_and_f_trans}
\beta=3.7\times10^{-9}\left(\frac{\gamma}{0.2}\right)^{-1/2}\left(\frac{g_{\star,\rm{form}}}{10.75}\right)^{1/4}\left(\frac{M_{\rm{PBH}}}{M_{\odot}}\right)f_{\rm{PBH}},
\end{equation}
where $g_{\star,\rm{form}}$ represents the relativistic degrees of freedom, which is taken to be $g_{\star,\rm{form}}=10.75$ in our case.

\subsection{CMB $\mu$ distortion and PBHs constraints} 

Although the CMB is very close to a black body spectrum with a temperature of 2.7 K, it still exhibits minor deviations.
These distortions can arise from several mechanisms, including reionization, structure formation, adiabatic cooling of baryons, cosmological recombination radiation, and the damping of small-scale perturbations. Each of these phenomena is closely tied to the cosmological evolution of the early Universe, providing valuable insights into the dynamics of cosmological perturbations as well \cite{chluba_which_2016,abitbol_prospects_2017,novikov_separation_2023}.

As the Universe expands and cools, different physical processes drive the formation of various CMB distortions. Before recombination, the principal component analysis method has shown that classifying these distortions into $\mu$-type and $y$-type is effective and sufficient, as any residual distortions, such as $r$-type, are relatively negligible. Specifically, at redshifts of $5 \times 10^4 \lesssim z \lesssim 2 \times 10^6$, only Compton scattering is efficient, prompting the photon distribution to evolve into a Bose-Einstein distribution presenting a $\mu$-type distortion characterized by the small chemical potential $\mu$ \cite{sunyaev_small_1970,hu_thermalization_1993,khatri_creation_2012}.
In this work, we are concerned about the dissipation of primordial perturbations caused by the diffusion of photons during the epoch. It is one of the most important distortion signals in the early universe models \cite{chluba_which_2016}.

Explaining the related physical process with the improved efficacy of PBH production commonly invokes an enhancement of the primordial power spectrum on small scales. 
The $\mu$-type distortion primarily constrains the damping tail of CMB anisotropies at $\ell \gtrsim 500$, providing crucial insights into the statistical properties of primordial perturbations at these smaller scales, including their non-trivial features and potential non-Gaussianity \cite{chluba_sunyaevzeldovich_2013,khatri_limits_2015,planck_collaboration_planck_2016,erler_plancks_2018,bianchini_cmb_2022}. While the magnitude of $\mu$ distortion is predicted to be small within the standard $\Lambda$CDM cosmology with $\mu \approx 2 \times 10^{-8}$ for the entire sky, it remains a critical observable for future experiments, such as the Primordial Inflation Explorer (PIXIE) \cite{hill_taking_2015,chluba_which_2016}.

The $\mu$ distortion coming from the energy injection earlier than recombination can be calculated as:  
\begin{equation} \label{eq:mu-dist}
    \mu \approx 1.401 \frac{\Delta \rho_{\gamma}}{\rho_{\gamma}}\bigg|_{\mu}=1.401\int^{\infty}_{z_{\mu}}\frac{\mathcal{J}(z)}{\rho_{\gamma}}\frac{\mathrm{d}Q}{\mathrm{d}z}\mathrm{d}z.
\end{equation}
 The visibility function denoted as $\mathcal{J}(z)$ quantifies the contributions to the observed $\mu$ distortion from a given energy injection across different redshift intervals. The specific form of $\mathcal{J}(z)$ can be derived through detailed consideration. But for a simple and effective analysis, we approximate $\mathcal{J}(z)$ as $\mathcal{J}(z)\approx \exp(-[z/z_{\mu}]^{5/2})$, where $z_{\mu} \approx 1.98 \times 10^6$. This simplification comes from the solely double Compton scattering contributions, which accurately approximate the full numerical simulations \cite{chluba_evolution_2012,chluba_cmb_2012,chluba_greens_2013}. It shows the $\mu$ distortion is fundamentally an integral over the effective energy release rate suppressed by the visibility function. The magnitude of energy injection can be described by the effective heating function $\frac{1}{\rho_{\gamma}}\frac{\mathrm{d}Q}{\mathrm{d}z}$. Considering the damping of small-scale perturbations as a source of $\mu$ distortion, the effective energy release rate is intricately linked to the primordial power spectrum as:
\begin{equation}
    \frac{1}{\rho_{\gamma}}\frac{\mathrm{d}Q}{\mathrm{d}z}=\frac{4\dot{\tau} \langle S \rangle }{H(1+z)},
\end{equation}
 where $\langle S \rangle$ represents the source term in the Boltzmann equation and $\tau$ represents the Thomson optical depth. Under the tight coupling approximation, this source term can be derived from small-scale perturbations \cite{chluba_cmb_2012}:
\begin{equation}  
 \langle S \rangle \approx \frac{\alpha_v}{\tau^{\prime}} \partial_\eta k_{\mathrm{D}}^{-2} \int \frac{\mathrm{d}^3 k}{(2 \pi)^3} k^2 P_\zeta(k) 2 \sin ^2\left(k r_{\mathrm{s}}\right) e^{-2 k^2 / k_{\mathrm{D}}^2}.
\end{equation}
In this source term, $\alpha_{\nu}\approx 0.81$ represents the contribution of neutrino to energy density; $\eta$  denotes the conformal time; $r_s$ means the sound horizon and $\tau'$ is the derivative of Thomson optical depth to conformal time; $k_{\mathrm{D}}$ is the damping scale which can be expressed as $k_\mathrm{D}\approx 4.0\times10^{-6}(1+z)^{3/2} \ \mathrm{Mpc}^{-1}$, and $P_{\zeta}(k)=2\pi^2 \mathcal{P}_\zeta(k)/k^3$ denotes the primordial power spectrum of curvature perturbations.

In the limit, $c_s^2\simeq1/3$, the effective energy release rate for the photon field is therefore given by
\begin{equation}\label{Eq:mu_calculation}
    \frac{1}{\rho_{\gamma}} \frac{\mathrm{d}Q}{\mathrm{d}z} \approx 9.4 a \int \frac{k \mathrm{~d} k}{k_{\mathrm{D}}^2} \mathcal{P}_\zeta(k) 2 \sin ^2\left(k r_{\mathrm{s}}\right) e^{-2 k^2 / k_{\mathrm{D}}^2}.
\end{equation}
Assuming a non-trivial deviation of the full power spectrum from a standard near-scale-invariant spectrum is represented by a Dirac-$\delta$ peak at the scale around $k_\delta$, the power spectrum can be expressed as
\begin{equation}
    \mathcal{P}^{\rm{delta}}_{\zeta}(k)=A_{\delta} k \cdot \delta_\mathrm{D}(k-k_{\delta}),
\end{equation}
where $A_{\delta}$ denotes the amplitude of the non-trivial features. Based on this spectrum, an approximation for the $\mu$ distortion due to damping can be given by \cite{chluba_probing_2012}:
\begin{equation} \label{Eq:MuDistDelta}
    \mu \approx 2.2 A_{\delta} \Bigg[ \mathrm{exp}\left(-\frac{\hat{k}_{\delta}}{5400}\right)-\mathrm{exp}\left(-\frac{\hat{k}^2_{\delta}}{31.6^2}\right) \Bigg],
\end{equation}
where $k=\hat{k} \ {\rm{Mpc}}^{-1}$. 
As the standard near-scale-invariant power spectrum contributes to $\mu$ distortion much less than observational constraints, we focus on the $\mu$ distortion produced by the non-trivial features of the full power spectrum only, as shown in Eq. \eqref{Eq:MuDistDelta} \cite{chluba_distinguishing_2013,kite_bridging_2021,10.1093/mnras/stad3861}. In most PBH formation scenarios, PBHs are formed from the overdense perturbations characterized by non-trivial features on the power spectrum. Therefore, limits on PBH abundance can be derived from the constraints on these non-trivial features as indicated by $\mu$-type distortion.

The analysis above reveals that the PBH abundance is tightly constrained within the mass range of $10^4 \sim 10^{11} M_{\odot}$ \cite{nakama_limits_2018}. This result reduces the possibility that supermassive black holes within this mass range could originate from PBHs. Furthermore, considering primordial non-Gaussianity might mitigate these constraints, potentially allowing for a higher PBH abundance within the specified mass range \cite{Hooper_2024}. Recent findings suggest that $\zeta$ may not be optimal for calculating primordial black hole abundance \cite{de_luca_note_2022}. Consequently, it is worthwhile to consider more refined approaches. Therefore, in the subsequent section, we will illustrate how to use alternative methodologies to compute the PBH abundance can provide a more accurate assessment.

\subsection{Compaction function analysis of PBH abundance}
The conventional method of calculating the abundance of primordial black holes based on curvature may pose risks because it takes into account the long-wavelength mode, which should not affect whether PBHs will form or not \cite{young_calculating_2014}.  A more reliable approach is to utilize local observables, such as density contrast or compaction function, to perform the calculation. Moreover, the latter is shown to exhibit a threshold that can be directly driven by an averaged compaction function, whose averaged threshold is robust when considering different profiles of perturbations \cite{shibata_black_1999,PhysRevD.91.084057,Escrivà_2021}. The compaction function is defined as
\begin{equation}
    \mathcal{C}=\frac{2[M(r,t)-M_\mathrm{b}(r,t)]}{R(r,t)},
\end{equation}
where $R$ is areal radius, $M=\int^R_04\pi \rho \tilde{R}^2 \mathrm{d}\tilde{R}$ is the Misner-Sharp-Hernandez mass and the subscript of $M_\mathrm{b}$ means the background. 

In calculating the threshold $\mathcal{C}_{\rm{th}}$, we considered the impact from the profile of perturbations. To evaluate the probability of threshold exceeding, a 2D joint probability distribution was employed \cite{gow_non-perturbative_2023}.
Additionally, PBH formed from critical collapse is also in our consideration \cite{choptuik_universality_1993,evans_critical_1994,niemeyer_near-critical_1998}. 
While our discussion is confined to the most straightforward profile and Gaussian distribution of random variables, the methodology remains adaptable to non-Gaussian cases or alternative profiles. For detailed calculations, please refer to Appendix.~\ref{PBHs_cal_details}.

Moreover, we aim to investigate the effects of changing the power spectrum from $\delta$ form to extended features. Therefore, two features are considered, they are more realistic among inflation models \cite{Pi_2023,PhysRevD.102.103527,meng_primordial_2023}. The first one is a lognormal (``ln'' in short) type power spectrum with one peak in the following form:
\begin{equation}
    \mathcal{P}^{\rm{ln}}_{\zeta}(k)=\frac{A_{\rm{ln}}}{\sqrt{2\pi}\sigma_\mathrm{ln}}\exp\Bigg[-\frac{1}{2}\left(\frac{\ln k- \ln k_\mathrm{p}}{\sigma_\mathrm{ln}}\right)^2\Bigg],
\end{equation}
where $k_\mathrm{p}$ and $\sigma_\mathrm{ln}$ are approximately the center and width of the peak, respectively. The second is a power spectrum peak in the box form, which can be expressed as follows:
\begin{equation}
\mathcal{P}^{\rm{box}}_{\zeta}(k)=A_{\rm{box}} \Theta(k-k_{\rm{min}})\Theta(k_{\rm{max}}-k),
\end{equation}
where $\Theta$ is the step function, we can also represent it with the position of the center of the step $k_{\rm{middle}}=(k_{\rm{min}}+k_{\rm{max}})/2$ and width $\Delta=k_{\rm{max}}-k_{\rm{min}}$.
We will use the FIRAS data to examine constraints of the PBH abundance given by the various non-trivial types of power spectrum from $\mu$ distortion.

\begin{figure}
    \centering
    \includegraphics[width=1.0\linewidth]{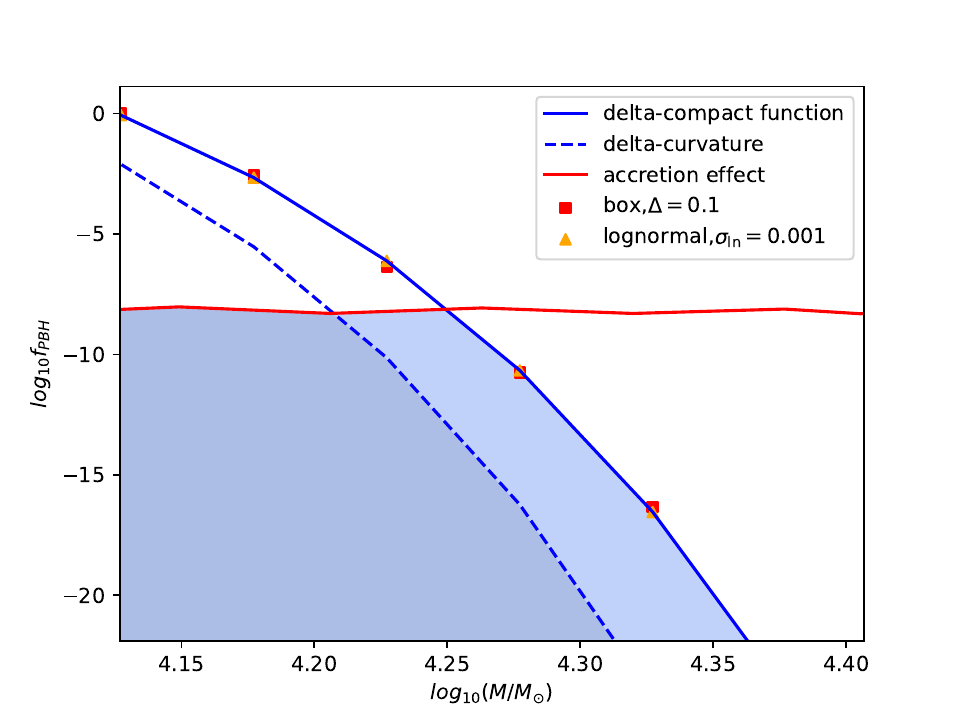}
    \caption{
    The improvement by our method to the PBH abundance constraints on the lower mass region we considered \cite{nakama_limits_2018}. $f_\mathrm{PBH}$ represents the fraction of PBHs against the total dark matter component and $M$ is given by the mass where the maximum of the mass function is.
}\label{fig1}

\end{figure}

\section{result and analysis}\label{sec:section3}

Based on the discussion, we will demonstrate how the compaction function modifies the constraints on PBH abundance from $\mu$ distortion, focusing on PBHs with masses in the range of $10^4\sim10^{14} M_{\odot}$.
For a given central position and width of a feature, the observational upper limits on $\mu$ distortion constrain the amplitude of the feature (e.g., $A_\delta$, $A_\mathrm{ln}$, $A_\mathrm{box}$), which in turn constrains PBH abundance as derived from the power spectrum. 

Figure.~\ref{fig1} illustrates the constraints on the abundance of PBHs on the lower mass region ($10^{4.1}\sim10^{4.4}M_\odot$) we considered. The blue curves represent the results with a delta-form featured power spectrum $\mathcal{P}^{\rm{delta}}_{\zeta}(k)$. The dashed curve indicates the upper limit on the possible PBH abundance as determined by the previous approach \cite{nakama_limits_2018}. The solid curve represents the permissible range of PBH abundance computed using the compaction function method, highlighting a relaxation of the limit. Specifically, the dark blue area represents the PBH abundance permitted with both methods, while the light blue area is only allowed using the compaction function. The red curve delineates the upper limit of PBH abundance ascertainable through CMB anisotropy observations, specifically focusing on the limits imposed by PBH accretion \cite{carr_constraints_2021}. For the mass region $M_{\rm{PBH}} \gtrsim 10^{4.25}M_{\odot}$ that is not constrained by PBH accretion, the potential abundance amplifies by approximately 5 orders of magnitude with compaction function method.
 The red and yellow points in the figure represent the limits on PBH abundance when considering the power spectrum with lognormal $\mathcal{P}^{\rm{ln}}_{\zeta}(k)$ and box $\mathcal{P}^{\rm{box}}_{\zeta}(k)$ form, respectively. The difference is about $1\% \sim 2\%$, which makes them indistinguishable concerning computational errors. Consequently, this suggests that variations on the features of the power spectrum contribute negligible effects on the results within the considered mass range and the width of the feature. It shows that the $\delta$ form power spectrum feature $\mathcal{P}^{\rm{delta}}_{\zeta}(k)$ serves as an adequate approximation for the extended power spectra with a relatively narrow peak.

\begin{figure}
    \centering
    \includegraphics[width=1.0\linewidth]{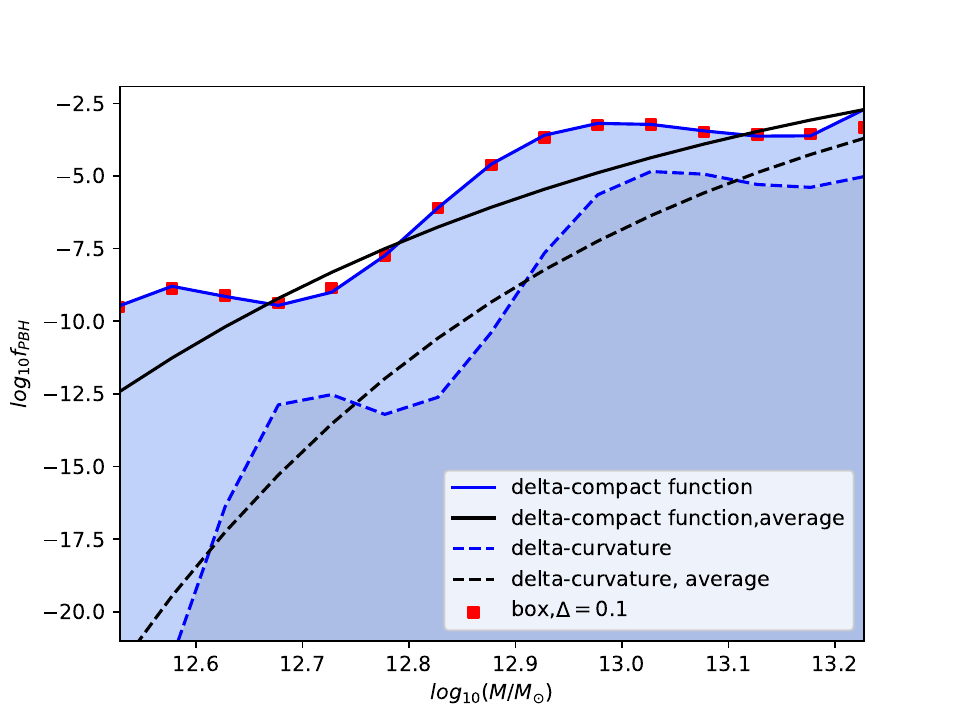}
     \caption{
    The improved PBH abundance constraints in the massive region and the deviation in the results given by the extended power spectrum feature. Without other constraints in this mass range ($10^{12.5}-10^{13.25} M_{\odot}$), our method increases the possible total abundance of PBHs constrained by $\mu$ distortion. As previously mentioned, the dark blue area shows the allowed abundance of PBHs in both methods while the light part shows the abundance only allowed by the new method. 
    }\label{fig2}
\end{figure}

Figure.~\ref{fig2} presents the constraints on the larger mass region ($10^{12.5}\sim10^{13.25} M_{\odot}$). Similar to the results on lower mass region, calculations utilizing the compaction function method indicate a possible amplification of the PBH abundance by approximately $1.5\sim10$ orders of magnitude. This significant amplification supports the cosmological models that produce the PBHs within this mass range to be viable candidates for SMBH seeds or to contribute to other physical processes. Notably, even with the traditional method, our results still exhibit slight oscillating deviation compared to those reported in the previous research \cite{nakama_limits_2018}. This deviation is primarily due to preserving the sin-squared term $\mathrm{sin}^2(kr)$ in $\mu$ distortion instead of averaging it as 0.5. The oscillation feature also recovers in the result by using the compaction function. We also present the averaged result in Figure.~\ref{fig2}, and it shows a similar amplification we obtained in precise calculations, implying the convincing of the approximation.
The red dots in Figure.~\ref{fig2} illustrate the impact of changing to the box-form feature in the power spectrum $\mathcal{P}^{\rm{box}}_{\zeta}(k)$ on the results, indicating corrections of approximately $1\%\sim2\%$. Given the constraints on the width $\Delta$ of the feature coming from the position of the peak $k_\mathrm{middle}$, it suggests that incorporating extended power spectra feature within the relevant frequency band $k$ does not result in significant adjustments to constraints on PBH abundance given by $\mathcal{P}^{\rm{delta}}_{\zeta}(k)$. This result emphasizes the robustness of the constraints concerning the variations of the spectral form within the considered large mass region when employing the compaction function approach.

\begin{figure}
\centering

\includegraphics[width=1.0\linewidth]{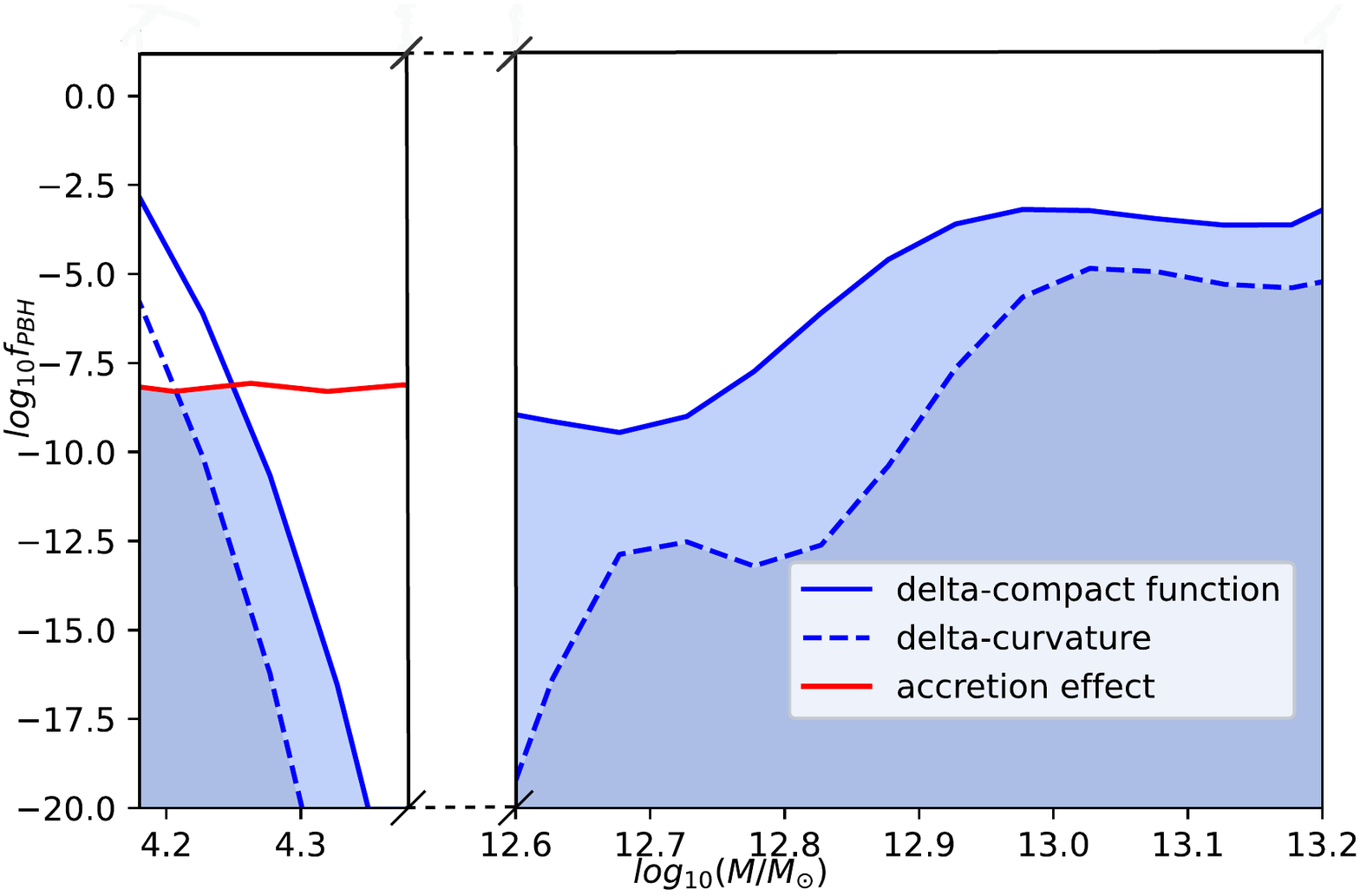}
    \caption{ The joint constraints given by both compaction function and curvature methods across the mass range $10^4 - 10^{14} M_\odot$. The dark blue area and the light part show the abundance allowed by both methods or only the new method. The red solid curve indicates the limit from accretion effects. They are indirectly constrained by observations from FIRAS $\mu < 9\times 10^{-5}$, and PIXIE will provide tighter limits in the future.  }\label{fig3}
\end{figure}

Figure.~\ref{fig3} summarizes the modifications introduced by the compaction function method, which significantly influences the constraints on the abundance of PBHs across the mass range. The changes are more evident in the massive region of the mass range we are concerned about. This approach slightly lets loose the $\mu$ distortion constraints on massive PBHs that were thought to be ruled out and suggests a broader scope of scenarios for the structure and composition of the Universe.

According to the results, we find that an extended primordial power spectrum (as demonstrated by $\mathcal{P}^{\rm{ln}}_{\zeta}(k)$ and $\mathcal{P}^{\rm{box}}_{\zeta}(k)$) does not yield tighter constraints on $f_{\rm PBH}$ as the previous work have revealed \cite{nakama_limits_2018,Hooper_2024}. Indeed, as seen directly from Eq.~(\ref{Eq:mu_calculation}), $\mathcal{P}_{\zeta}(k)$ always contributes positively to the integral that sources $\mu$ distortion. Therefore, the amplitude of an extended $\mathcal{P}_{\zeta}(k)$ is suppressed by a given $\mu$ distortion measurement, such as $\mu < 9 \times 10^{-5}$ observed by FIRAS. Consequently, the abundance of PBHs corresponding to the peak scale ($k_{\rm p}$ or $k_{\rm middle}$) is also suppressed (see Eq.~(\ref{eq:PBHbeta})). However, an extended $\mathcal{P}_{\zeta}(k)$ produces PBHs with a broader mass function. So, when calculating $f_{\rm PBH}$, it is essential to consider the effects of a non-monochromatic PBH mass function. In this Letter, we integrated the PBH mass function over the mass to obtain $f_{\rm PBH}$ and mapped it to $M_{\rm PBH}$ corresponding to the peak scale in Figure.~\ref{fig1} and Figure.~\ref{fig2}. Our analysis shows that within the frequency bands $k < 10^{0.3}\ \rm{Mpc}^{-1}$ and $k > 10^{4.5}\ \rm{Mpc}^{-1}$ (corresponding to mass ranges $M_{\rm PBH} < 10^{4.23} M_{\odot}$ and $M_{\rm PBH} > 10^{12.6} M_{\odot}$), a narrowly extended $\mathcal{P}_{\zeta}(k)$ with different shapes does not significantly affect the constraints.
 
\section{conclusion}\label{sec:section4}

As an important candidate for 
the seeds of SMBHs, the constraints on the possible abundance of PBHs have garnered significant interest in cosmology. These constraints and the possible detection of PBHs in the future will yield valuable insights into the related physical process and cosmological models. Meanwhile, with the development of the observations of CMB, it is possible to gain a precise magnitude of CMB distortions, which can provide tight constraints on the primordial power spectrum and the abundance of massive PBH. This research has extended our knowledge of the constraints on PBH abundance within the related mass range ($10^4\sim10^{14} M_{\odot}$). 

Recent studies on PBHs have revealed that even if the curvature perturbation $\zeta$ follows a Gaussian distribution, the corresponding density fluctuations in real space can exhibit non-Gaussian characteristics. Previous research predominantly employed traditional curvature-based methods to calculate PBH abundance from the primordial power spectrum. However, in this work, we adopt a methodology based on the compaction function to refine the calculation of PBH abundance, aiming to provide more accurate constraints from CMB $\mu$ distortion. Additionally, we account for other factors that might influence the results, such as variations in the perturbation profile and modifications to the threshold criteria, as well as a two-dimensional perturbation probability distribution to address the likelihood of PBH formation \cite{musco_threshold_2019,gow_non-perturbative_2023}. Our findings indicate that even in the absence of non-Gaussianity in $\zeta$, this refined methodology significantly alleviates previous constraints on PBH abundance from CMB $\mu$ distortion. Specifically, in the lower mass range ($10^4 \sim 10^{4.25} M_{\odot}$), the constraints have been relaxed by approximately 5 orders of magnitude, while in the more massive range ($10^{12.5} \sim 10^{13} M_{\odot}$), they have been loosened by $1.5 \sim 10$ orders of magnitude. These results suggest that early universe models capable of producing massive PBHs within the considered mass range remain viable under current CMB distortion constraints. Since the maximum enhancement of the power spectrum is still limited by CMB distortions, further relaxation of PBH abundance constraints may require the introduction of other mechanisms, such as non-Gaussian perturbations \cite{nakama_limits_2018, Hooper_2024}.

We also analyze the impact of various forms of extended features on the power spectrum and the constraints in detail. It shows that the constraints on PBH abundance remain almost the same to $\mathcal{P}^{\rm{delta}}_{\zeta}(k)$ in $1\%\sim2\%$ when considering lognormal $\mathcal{P}^{\rm{ln}}_{\zeta}(k)$ or box $\mathcal{P}^{\rm{box}}_{\zeta}(k)$ feature with narrow width. Thus, the delta-form power spectrum $\mathcal{P}^{\rm{delta}}_{\zeta}(k)$ is an appropriate approximation for the realistic power spectrum with a narrow peak.

This work inspires the constraint of PBH abundance and CMB distortion. Firstly, the methodology can be extended to $y$ distortion, possibly resulting from damping primordial perturbations  \cite{10.1093/mnras/stad3861}. Then, our methodology is adaptable to cosmological models featuring non-Gaussian perturbations by directly adjusting the two-dimensional probability distribution functions. In this way, combining our methods and the perturbations beyond Gaussian, we may find models that are valid under the robust CMB distortion produce PBHs contribute to the SMBH seeds \cite{byrnes_robust_2024,Hooper_2024,nakama_limits_2018}. Moreover, as we assumed the extended feature on the power spectrum with a narrow width, the wide bump on the power spectrum will naturally be considered. CMB distortion will suppress its amplitude regarding the wide bump on the power spectrum. However, a more comprehensive investigation is necessary to furnish further details regarding PBH abundance within the massive mass range \cite{byrnes_robust_2024}.
Additionally, these results can be extended to forthcoming observations from PIXIE, enriching our comprehension of primordial perturbations and, thereby the physics of PBHs \cite{a_kogut_primordial_2011,dent_cosmological_2012,abitbol_prospects_2017}. This investigation advances our theoretical understanding and observational concerning PBH dynamics.

\section*{Acknowledgements}
We thank Misao Sasaki, Masahide Yamaguchi, Cheng-Feng Tang, Bo Wang and Geyu Mo for their valuable comments. 
This work is supported in part by National Key R\&D Program of China (2021YFC2203100), by NSFC (12261131497), by CAS young interdisciplinary innovation team (JCTD-2022-20), by 111 Project (B23042), by Fundamental Research Funds for Central Universities, by the postdoctoral fellowship program of CPSF under grant number GZC20240212, by CSC Innovation Talent Funds, by USTC Fellowship for International Cooperation, by USTC Research Funds of the Double First-Class Initiative. 
Kavli IPMU is supported by World Premier International Research Center Initiative (WPI), MEXT, Japan. We acknowledge the use of the computing facilities of TIT, as well as the clusters LINDA and JUDY of the particle cosmology group at USTC.

\bibliography{main} 

\begin{thebibliography}{103}%
\makeatletter
\providecommand \@ifxundefined [1]{%
 \@ifx{#1\undefined}
}%
\providecommand \@ifnum [1]{%
 \ifnum #1\expandafter \@firstoftwo
 \else \expandafter \@secondoftwo
 \fi
}%
\providecommand \@ifx [1]{%
 \ifx #1\expandafter \@firstoftwo
 \else \expandafter \@secondoftwo
 \fi
}%
\providecommand \natexlab [1]{#1}%
\providecommand \enquote  [1]{``#1''}%
\providecommand \bibnamefont  [1]{#1}%
\providecommand \bibfnamefont [1]{#1}%
\providecommand \citenamefont [1]{#1}%
\providecommand \href@noop [0]{\@secondoftwo}%
\providecommand \href [0]{\begingroup \@sanitize@url \@href}%
\providecommand \@href[1]{\@@startlink{#1}\@@href}%
\providecommand \@@href[1]{\endgroup#1\@@endlink}%
\providecommand \@sanitize@url [0]{\catcode `\\12\catcode `\$12\catcode `\&12\catcode `\#12\catcode `\^12\catcode `\_12\catcode `\%12\relax}%
\providecommand \@@startlink[1]{}%
\providecommand \@@endlink[0]{}%
\providecommand \url  [0]{\begingroup\@sanitize@url \@url }%
\providecommand \@url [1]{\endgroup\@href {#1}{\urlprefix }}%
\providecommand \urlprefix  [0]{URL }%
\providecommand \Eprint [0]{\href }%
\providecommand \doibase [0]{http://dx.doi.org/}%
\providecommand \selectlanguage [0]{\@gobble}%
\providecommand \bibinfo  [0]{\@secondoftwo}%
\providecommand \bibfield  [0]{\@secondoftwo}%
\providecommand \translation [1]{[#1]}%
\providecommand \BibitemOpen [0]{}%
\providecommand \bibitemStop [0]{}%
\providecommand \bibitemNoStop [0]{.\EOS\space}%
\providecommand \EOS [0]{\spacefactor3000\relax}%
\providecommand \BibitemShut  [1]{\csname bibitem#1\endcsname}%
\let\auto@bib@innerbib\@empty
%</preamble>
\bibitem [{\citenamefont {Zel'dovich}\ and\ \citenamefont {Novikov}()}]{Zeldovich:1967lct}%
  \BibitemOpen
  \bibfield  {author} {\bibinfo {author} {\bibfnamefont {Y.~B.}\ \bibnamefont {Zel'dovich}}\ and\ \bibinfo {author} {\bibfnamefont {I.~D.}\ \bibnamefont {Novikov}},\ }\href@noop {} {\bibfield  {journal} {\bibinfo  {journal} {Sov. Astron.}\ }\textbf {\bibinfo {volume} {10}},\ \bibinfo {pages} {602}}\BibitemShut {NoStop}%
\bibitem [{\citenamefont {Hawking}(1971)}]{hawking_gravitationally_1971}%
  \BibitemOpen
  \bibfield  {author} {\bibinfo {author} {\bibfnamefont {S.}~\bibnamefont {Hawking}},\ }\href {\doibase 10.1093/mnras/152.1.75} {\bibfield  {journal} {\bibinfo  {journal} {Mon. Not. R. Astron. Soc.}\ }\textbf {\bibinfo {volume} {152}},\ \bibinfo {pages} {75} (\bibinfo {year} {1971})}\BibitemShut {NoStop}%
\bibitem [{\citenamefont {Carr}\ and\ \citenamefont {Hawking}(1974)}]{carr_black_1974}%
  \BibitemOpen
  \bibfield  {author} {\bibinfo {author} {\bibfnamefont {B.~J.}\ \bibnamefont {Carr}}\ and\ \bibinfo {author} {\bibfnamefont {S.~W.}\ \bibnamefont {Hawking}},\ }\href {\doibase 10.1093/mnras/168.2.399} {\bibfield  {journal} {\bibinfo  {journal} {Mon. Not. R. Astron. Soc.}\ }\textbf {\bibinfo {volume} {168}},\ \bibinfo {pages} {399} (\bibinfo {year} {1974})}\BibitemShut {NoStop}%
\bibitem [{\citenamefont {Carr}(1975)}]{carr1975primordial}%
  \BibitemOpen
  \bibfield  {author} {\bibinfo {author} {\bibfnamefont {B.~J.}\ \bibnamefont {Carr}},\ }\href@noop {} {\enquote {\bibinfo {title} {The primordial black hole mass spectrum},}\ } (\bibinfo {year} {1975})\BibitemShut {NoStop}%
\bibitem [{\citenamefont {Khlopov}(2010)}]{khlopov_primordial_2010}%
  \BibitemOpen
  \bibfield  {author} {\bibinfo {author} {\bibfnamefont {M.~Y.}\ \bibnamefont {Khlopov}},\ }\href {\doibase 10.1088/1674-4527/10/6/001} {\bibfield  {journal} {\bibinfo  {journal} {Res. Astron. Astrophys.}\ }\textbf {\bibinfo {volume} {10}},\ \bibinfo {pages} {495} (\bibinfo {year} {2010})}\BibitemShut {NoStop}%
\bibitem [{\citenamefont {Chapline}(1975)}]{Chapline:1975ojl}%
  \BibitemOpen
  \bibfield  {author} {\bibinfo {author} {\bibfnamefont {G.~F.}\ \bibnamefont {Chapline}},\ }\href {\doibase 10.1038/253251a0} {\bibfield  {journal} {\bibinfo  {journal} {Nature}\ }\textbf {\bibinfo {volume} {253}},\ \bibinfo {pages} {251} (\bibinfo {year} {1975})}\BibitemShut {NoStop}%
\bibitem [{\citenamefont {Alcock}\ \emph {et~al.}(1997)\citenamefont {Alcock} \emph {et~al.}}]{MACHO:1996qam}%
  \BibitemOpen
  \bibfield  {author} {\bibinfo {author} {\bibfnamefont {C.}~\bibnamefont {Alcock}} \emph {et~al.} (\bibinfo {collaboration} {MACHO}),\ }\href {\doibase 10.1086/304535} {\bibfield  {journal} {\bibinfo  {journal} {Astrophys. J.}\ }\textbf {\bibinfo {volume} {486}},\ \bibinfo {pages} {697} (\bibinfo {year} {1997})}\BibitemShut {NoStop}%
\bibitem [{\citenamefont {Bird}\ \emph {et~al.}(2016)\citenamefont {Bird}, \citenamefont {Cholis}, \citenamefont {Mu\~noz}, \citenamefont {Ali-Ha\"\i{}moud}, \citenamefont {Kamionkowski}, \citenamefont {Kovetz}, \citenamefont {Raccanelli},\ and\ \citenamefont {Riess}}]{Bird:2016dcv}%
  \BibitemOpen
  \bibfield  {author} {\bibinfo {author} {\bibfnamefont {S.}~\bibnamefont {Bird}}, \bibinfo {author} {\bibfnamefont {I.}~\bibnamefont {Cholis}}, \bibinfo {author} {\bibfnamefont {J.~B.}\ \bibnamefont {Mu\~noz}}, \bibinfo {author} {\bibfnamefont {Y.}~\bibnamefont {Ali-Ha\"\i{}moud}}, \bibinfo {author} {\bibfnamefont {M.}~\bibnamefont {Kamionkowski}}, \bibinfo {author} {\bibfnamefont {E.~D.}\ \bibnamefont {Kovetz}}, \bibinfo {author} {\bibfnamefont {A.}~\bibnamefont {Raccanelli}}, \ and\ \bibinfo {author} {\bibfnamefont {A.~G.}\ \bibnamefont {Riess}},\ }\href {\doibase 10.1103/PhysRevLett.116.201301} {\bibfield  {journal} {\bibinfo  {journal} {Phys. Rev. Lett.}\ }\textbf {\bibinfo {volume} {116}},\ \bibinfo {pages} {201301} (\bibinfo {year} {2016})}\BibitemShut {NoStop}%
\bibitem [{\citenamefont {Hall}\ \emph {et~al.}(2020)\citenamefont {Hall}, \citenamefont {Gow},\ and\ \citenamefont {Byrnes}}]{Hall:2020daa}%
  \BibitemOpen
  \bibfield  {author} {\bibinfo {author} {\bibfnamefont {A.}~\bibnamefont {Hall}}, \bibinfo {author} {\bibfnamefont {A.~D.}\ \bibnamefont {Gow}}, \ and\ \bibinfo {author} {\bibfnamefont {C.~T.}\ \bibnamefont {Byrnes}},\ }\href {\doibase 10.1103/PhysRevD.102.123524} {\bibfield  {journal} {\bibinfo  {journal} {Phys. Rev. D}\ }\textbf {\bibinfo {volume} {102}},\ \bibinfo {pages} {123524} (\bibinfo {year} {2020})}\BibitemShut {NoStop}%
\bibitem [{\citenamefont {De~Luca}\ \emph {et~al.}(2020)\citenamefont {De~Luca}, \citenamefont {Franciolini}, \citenamefont {Pani},\ and\ \citenamefont {Riotto}}]{DeLuca:2020fpg}%
  \BibitemOpen
  \bibfield  {author} {\bibinfo {author} {\bibfnamefont {V.}~\bibnamefont {De~Luca}}, \bibinfo {author} {\bibfnamefont {G.}~\bibnamefont {Franciolini}}, \bibinfo {author} {\bibfnamefont {P.}~\bibnamefont {Pani}}, \ and\ \bibinfo {author} {\bibfnamefont {A.}~\bibnamefont {Riotto}},\ }\href {\doibase 10.1103/PhysRevD.102.043505} {\bibfield  {journal} {\bibinfo  {journal} {Phys. Rev. D}\ }\textbf {\bibinfo {volume} {102}},\ \bibinfo {pages} {043505} (\bibinfo {year} {2020})}\BibitemShut {NoStop}%
\bibitem [{\citenamefont {Franciolini}\ \emph {et~al.}(2022)\citenamefont {Franciolini}, \citenamefont {Baibhav}, \citenamefont {De~Luca}, \citenamefont {Ng}, \citenamefont {Wong}, \citenamefont {Berti}, \citenamefont {Pani}, \citenamefont {Riotto},\ and\ \citenamefont {Vitale}}]{Franciolini:2021tla}%
  \BibitemOpen
  \bibfield  {author} {\bibinfo {author} {\bibfnamefont {G.}~\bibnamefont {Franciolini}}, \bibinfo {author} {\bibfnamefont {V.}~\bibnamefont {Baibhav}}, \bibinfo {author} {\bibfnamefont {V.}~\bibnamefont {De~Luca}}, \bibinfo {author} {\bibfnamefont {K.~K.~Y.}\ \bibnamefont {Ng}}, \bibinfo {author} {\bibfnamefont {K.~W.~K.}\ \bibnamefont {Wong}}, \bibinfo {author} {\bibfnamefont {E.}~\bibnamefont {Berti}}, \bibinfo {author} {\bibfnamefont {P.}~\bibnamefont {Pani}}, \bibinfo {author} {\bibfnamefont {A.}~\bibnamefont {Riotto}}, \ and\ \bibinfo {author} {\bibfnamefont {S.}~\bibnamefont {Vitale}},\ }\href {\doibase 10.1103/PhysRevD.105.083526} {\bibfield  {journal} {\bibinfo  {journal} {Phys. Rev. D}\ }\textbf {\bibinfo {volume} {105}},\ \bibinfo {pages} {083526} (\bibinfo {year} {2022})}\BibitemShut {NoStop}%
\bibitem [{\citenamefont {Abbott}\ \emph {et~al.}(2022)\citenamefont {Abbott} \emph {et~al.}}]{LIGOScientific:2021job}%
  \BibitemOpen
  \bibfield  {author} {\bibinfo {author} {\bibfnamefont {R.}~\bibnamefont {Abbott}} \emph {et~al.} (\bibinfo {collaboration} {LIGO Scientific, VIRGO, KAGRA}),\ }\href {\doibase 10.1103/PhysRevLett.129.061104} {\bibfield  {journal} {\bibinfo  {journal} {Phys. Rev. Lett.}\ }\textbf {\bibinfo {volume} {129}},\ \bibinfo {pages} {061104} (\bibinfo {year} {2022})}\BibitemShut {NoStop}%
\bibitem [{\citenamefont {Afzal}\ \emph {et~al.}(2023)\citenamefont {Afzal} \emph {et~al.}}]{NANOGrav:2023hvm}%
  \BibitemOpen
  \bibfield  {author} {\bibinfo {author} {\bibfnamefont {A.}~\bibnamefont {Afzal}} \emph {et~al.} (\bibinfo {collaboration} {NANOGrav}),\ }\href {\doibase 10.3847/2041-8213/acdc91} {\bibfield  {journal} {\bibinfo  {journal} {Astrophys. J. Lett.}\ }\textbf {\bibinfo {volume} {951}},\ \bibinfo {pages} {L11} (\bibinfo {year} {2023})}\BibitemShut {NoStop}%
\bibitem [{\citenamefont {Chen}\ \emph {et~al.}(2022)\citenamefont {Chen}, \citenamefont {Yuan},\ and\ \citenamefont {Huang}}]{Chen:2021nxo}%
  \BibitemOpen
  \bibfield  {author} {\bibinfo {author} {\bibfnamefont {Z.-C.}\ \bibnamefont {Chen}}, \bibinfo {author} {\bibfnamefont {C.}~\bibnamefont {Yuan}}, \ and\ \bibinfo {author} {\bibfnamefont {Q.-G.}\ \bibnamefont {Huang}},\ }\href {\doibase 10.1016/j.physletb.2022.137040} {\bibfield  {journal} {\bibinfo  {journal} {Phys. Lett. B}\ }\textbf {\bibinfo {volume} {829}},\ \bibinfo {pages} {137040} (\bibinfo {year} {2022})}\BibitemShut {NoStop}%
\bibitem [{\citenamefont {Cai}\ \emph {et~al.}(2021{\natexlab{a}})\citenamefont {Cai}, \citenamefont {Jiang}, \citenamefont {Sasaki}, \citenamefont {Vardanyan},\ and\ \citenamefont {Zhou}}]{Cai:2021yvq}%
  \BibitemOpen
  \bibfield  {author} {\bibinfo {author} {\bibfnamefont {Y.-F.}\ \bibnamefont {Cai}}, \bibinfo {author} {\bibfnamefont {J.}~\bibnamefont {Jiang}}, \bibinfo {author} {\bibfnamefont {M.}~\bibnamefont {Sasaki}}, \bibinfo {author} {\bibfnamefont {V.}~\bibnamefont {Vardanyan}}, \ and\ \bibinfo {author} {\bibfnamefont {Z.}~\bibnamefont {Zhou}},\ }\href {\doibase 10.1103/PhysRevLett.127.251301} {\bibfield  {journal} {\bibinfo  {journal} {Phys. Rev. Lett.}\ }\textbf {\bibinfo {volume} {127}},\ \bibinfo {pages} {251301} (\bibinfo {year} {2021}{\natexlab{a}})}\BibitemShut {NoStop}%
\bibitem [{\citenamefont {Cai}\ \emph {et~al.}(2023{\natexlab{a}})\citenamefont {Cai}, \citenamefont {Chen}, \citenamefont {Wang},\ and\ \citenamefont {Yang}}]{Cai:2022kbp}%
  \BibitemOpen
  \bibfield  {author} {\bibinfo {author} {\bibfnamefont {R.-G.}\ \bibnamefont {Cai}}, \bibinfo {author} {\bibfnamefont {T.}~\bibnamefont {Chen}}, \bibinfo {author} {\bibfnamefont {S.-J.}\ \bibnamefont {Wang}}, \ and\ \bibinfo {author} {\bibfnamefont {X.-Y.}\ \bibnamefont {Yang}},\ }\href {\doibase 10.1088/1475-7516/2023/03/043} {\bibfield  {journal} {\bibinfo  {journal} {JCAP}\ }\textbf {\bibinfo {volume} {03}},\ \bibinfo {pages} {043} (\bibinfo {year} {2023}{\natexlab{a}})}\BibitemShut {NoStop}%
\bibitem [{\citenamefont {Cai}\ \emph {et~al.}(2023{\natexlab{b}})\citenamefont {Cai}, \citenamefont {He}, \citenamefont {Ma}, \citenamefont {Yan},\ and\ \citenamefont {Yuan}}]{Cai:2023dls}%
  \BibitemOpen
  \bibfield  {author} {\bibinfo {author} {\bibfnamefont {Y.-F.}\ \bibnamefont {Cai}}, \bibinfo {author} {\bibfnamefont {X.-C.}\ \bibnamefont {He}}, \bibinfo {author} {\bibfnamefont {X.-H.}\ \bibnamefont {Ma}}, \bibinfo {author} {\bibfnamefont {S.-F.}\ \bibnamefont {Yan}}, \ and\ \bibinfo {author} {\bibfnamefont {G.-W.}\ \bibnamefont {Yuan}},\ }\href {\doibase 10.1016/j.scib.2023.10.027} {\bibfield  {journal} {\bibinfo  {journal} {Sci. Bull.}\ }\textbf {\bibinfo {volume} {68}},\ \bibinfo {pages} {2929} (\bibinfo {year} {2023}{\natexlab{b}})}\BibitemShut {NoStop}%
\bibitem [{\citenamefont {Byrnes}\ \emph {et~al.}(2019)\citenamefont {Byrnes}, \citenamefont {Cole},\ and\ \citenamefont {Patil}}]{Byrnes:2018txb}%
  \BibitemOpen
  \bibfield  {author} {\bibinfo {author} {\bibfnamefont {C.~T.}\ \bibnamefont {Byrnes}}, \bibinfo {author} {\bibfnamefont {P.~S.}\ \bibnamefont {Cole}}, \ and\ \bibinfo {author} {\bibfnamefont {S.~P.}\ \bibnamefont {Patil}},\ }\href {\doibase 10.1088/1475-7516/2019/06/028} {\bibfield  {journal} {\bibinfo  {journal} {JCAP}\ }\textbf {\bibinfo {volume} {06}},\ \bibinfo {pages} {028} (\bibinfo {year} {2019})}\BibitemShut {NoStop}%
\bibitem [{\citenamefont {Pi}\ and\ \citenamefont {Sasaki}(2023)}]{Pi:2022ysn}%
  \BibitemOpen
  \bibfield  {author} {\bibinfo {author} {\bibfnamefont {S.}~\bibnamefont {Pi}}\ and\ \bibinfo {author} {\bibfnamefont {M.}~\bibnamefont {Sasaki}},\ }\href {\doibase 10.1103/PhysRevLett.131.011002} {\bibfield  {journal} {\bibinfo  {journal} {Phys. Rev. Lett.}\ }\textbf {\bibinfo {volume} {131}},\ \bibinfo {pages} {011002} (\bibinfo {year} {2023})}\BibitemShut {NoStop}%
\bibitem [{\citenamefont {Harada}\ \emph {et~al.}(2013)\citenamefont {Harada}, \citenamefont {Yoo},\ and\ \citenamefont {Kohri}}]{Harada:2013epa}%
  \BibitemOpen
  \bibfield  {author} {\bibinfo {author} {\bibfnamefont {T.}~\bibnamefont {Harada}}, \bibinfo {author} {\bibfnamefont {C.-M.}\ \bibnamefont {Yoo}}, \ and\ \bibinfo {author} {\bibfnamefont {K.}~\bibnamefont {Kohri}},\ }\href {\doibase 10.1103/PhysRevD.88.084051} {\bibfield  {journal} {\bibinfo  {journal} {Phys. Rev. D}\ }\textbf {\bibinfo {volume} {88}},\ \bibinfo {pages} {084051} (\bibinfo {year} {2013})}\BibitemShut {NoStop}%
\bibitem [{\citenamefont {Carr}\ \emph {et~al.}(2024)\citenamefont {Carr}, \citenamefont {Clesse}, \citenamefont {Garcia-Bellido}, \citenamefont {Hawkins},\ and\ \citenamefont {Kuhnel}}]{Carr:2023tpt}%
  \BibitemOpen
  \bibfield  {author} {\bibinfo {author} {\bibfnamefont {B.}~\bibnamefont {Carr}}, \bibinfo {author} {\bibfnamefont {S.}~\bibnamefont {Clesse}}, \bibinfo {author} {\bibfnamefont {J.}~\bibnamefont {Garcia-Bellido}}, \bibinfo {author} {\bibfnamefont {M.}~\bibnamefont {Hawkins}}, \ and\ \bibinfo {author} {\bibfnamefont {F.}~\bibnamefont {Kuhnel}},\ }\href {\doibase 10.1016/j.physrep.2023.11.005} {\bibfield  {journal} {\bibinfo  {journal} {Phys. Rept.}\ }\textbf {\bibinfo {volume} {1054}},\ \bibinfo {pages} {1} (\bibinfo {year} {2024})}\BibitemShut {NoStop}%
\bibitem [{\citenamefont {Escrivà}\ \emph {et~al.}(2024)\citenamefont {Escrivà}, \citenamefont {Kühnel},\ and\ \citenamefont {Tada}}]{Escriva:2022duf}%
  \BibitemOpen
  \bibfield  {author} {\bibinfo {author} {\bibfnamefont {A.}~\bibnamefont {Escrivà}}, \bibinfo {author} {\bibfnamefont {F.}~\bibnamefont {Kühnel}}, \ and\ \bibinfo {author} {\bibfnamefont {Y.}~\bibnamefont {Tada}},\ }in\ \href {\doibase https://doi.org/10.1016/B978-0-32-395636-9.00012-8} {\emph {\bibinfo {booktitle} {Black Holes in the Era of Gravitational-Wave Astronomy}}},\ \bibinfo {editor} {edited by\ \bibinfo {editor} {\bibfnamefont {M.~A.}\ \bibnamefont {Sedda}}, \bibinfo {editor} {\bibfnamefont {E.}~\bibnamefont {Bortolas}}, \ and\ \bibinfo {editor} {\bibfnamefont {M.}~\bibnamefont {Spera}}}\ (\bibinfo  {publisher} {Elsevier},\ \bibinfo {year} {2024})\ pp.\ \bibinfo {pages} {261--377}\BibitemShut {NoStop}%
\bibitem [{\citenamefont {Ivanov}\ \emph {et~al.}(1994)\citenamefont {Ivanov}, \citenamefont {Naselsky},\ and\ \citenamefont {Novikov}}]{Ivanov:1994pa}%
  \BibitemOpen
  \bibfield  {author} {\bibinfo {author} {\bibfnamefont {P.}~\bibnamefont {Ivanov}}, \bibinfo {author} {\bibfnamefont {P.}~\bibnamefont {Naselsky}}, \ and\ \bibinfo {author} {\bibfnamefont {I.}~\bibnamefont {Novikov}},\ }\href {\doibase 10.1103/PhysRevD.50.7173} {\bibfield  {journal} {\bibinfo  {journal} {Phys. Rev. D}\ }\textbf {\bibinfo {volume} {50}},\ \bibinfo {pages} {7173} (\bibinfo {year} {1994})}\BibitemShut {NoStop}%
\bibitem [{\citenamefont {Garcia-Bellido}\ \emph {et~al.}(1996)\citenamefont {Garcia-Bellido}, \citenamefont {Linde},\ and\ \citenamefont {Wands}}]{Garcia-Bellido:1996mdl}%
  \BibitemOpen
  \bibfield  {author} {\bibinfo {author} {\bibfnamefont {J.}~\bibnamefont {Garcia-Bellido}}, \bibinfo {author} {\bibfnamefont {A.~D.}\ \bibnamefont {Linde}}, \ and\ \bibinfo {author} {\bibfnamefont {D.}~\bibnamefont {Wands}},\ }\href {\doibase 10.1103/PhysRevD.54.6040} {\bibfield  {journal} {\bibinfo  {journal} {Phys. Rev. D}\ }\textbf {\bibinfo {volume} {54}},\ \bibinfo {pages} {6040} (\bibinfo {year} {1996})}\BibitemShut {NoStop}%
\bibitem [{\citenamefont {Quintin}\ and\ \citenamefont {Brandenberger}(2016)}]{Quintin:2016qro}%
  \BibitemOpen
  \bibfield  {author} {\bibinfo {author} {\bibfnamefont {J.}~\bibnamefont {Quintin}}\ and\ \bibinfo {author} {\bibfnamefont {R.~H.}\ \bibnamefont {Brandenberger}},\ }\href {\doibase 10.1088/1475-7516/2016/11/029} {\bibfield  {journal} {\bibinfo  {journal} {JCAP}\ }\textbf {\bibinfo {volume} {11}},\ \bibinfo {pages} {029} (\bibinfo {year} {2016})}\BibitemShut {NoStop}%
\bibitem [{\citenamefont {Hertzberg}\ and\ \citenamefont {Yamada}(2018)}]{Hertzberg:2017dkh}%
  \BibitemOpen
  \bibfield  {author} {\bibinfo {author} {\bibfnamefont {M.~P.}\ \bibnamefont {Hertzberg}}\ and\ \bibinfo {author} {\bibfnamefont {M.}~\bibnamefont {Yamada}},\ }\href {\doibase 10.1103/PhysRevD.97.083509} {\bibfield  {journal} {\bibinfo  {journal} {Phys. Rev. D}\ }\textbf {\bibinfo {volume} {97}},\ \bibinfo {pages} {083509} (\bibinfo {year} {2018})}\BibitemShut {NoStop}%
\bibitem [{\citenamefont {Ballesteros}\ and\ \citenamefont {Taoso}(2018)}]{Ballesteros:2017fsr}%
  \BibitemOpen
  \bibfield  {author} {\bibinfo {author} {\bibfnamefont {G.}~\bibnamefont {Ballesteros}}\ and\ \bibinfo {author} {\bibfnamefont {M.}~\bibnamefont {Taoso}},\ }\href {\doibase 10.1103/PhysRevD.97.023501} {\bibfield  {journal} {\bibinfo  {journal} {Phys. Rev. D}\ }\textbf {\bibinfo {volume} {97}},\ \bibinfo {pages} {023501} (\bibinfo {year} {2018})}\BibitemShut {NoStop}%
\bibitem [{\citenamefont {Franciolini}\ \emph {et~al.}(2018)\citenamefont {Franciolini}, \citenamefont {Kehagias}, \citenamefont {Matarrese},\ and\ \citenamefont {Riotto}}]{Franciolini:2018vbk}%
  \BibitemOpen
  \bibfield  {author} {\bibinfo {author} {\bibfnamefont {G.}~\bibnamefont {Franciolini}}, \bibinfo {author} {\bibfnamefont {A.}~\bibnamefont {Kehagias}}, \bibinfo {author} {\bibfnamefont {S.}~\bibnamefont {Matarrese}}, \ and\ \bibinfo {author} {\bibfnamefont {A.}~\bibnamefont {Riotto}},\ }\href {\doibase 10.1088/1475-7516/2018/03/016} {\bibfield  {journal} {\bibinfo  {journal} {JCAP}\ }\textbf {\bibinfo {volume} {03}},\ \bibinfo {pages} {016} (\bibinfo {year} {2018})}\BibitemShut {NoStop}%
\bibitem [{\citenamefont {Pi}\ \emph {et~al.}(2018)\citenamefont {Pi}, \citenamefont {Zhang}, \citenamefont {Huang},\ and\ \citenamefont {Sasaki}}]{Pi:2017gih}%
  \BibitemOpen
  \bibfield  {author} {\bibinfo {author} {\bibfnamefont {S.}~\bibnamefont {Pi}}, \bibinfo {author} {\bibfnamefont {Y.-l.}\ \bibnamefont {Zhang}}, \bibinfo {author} {\bibfnamefont {Q.-G.}\ \bibnamefont {Huang}}, \ and\ \bibinfo {author} {\bibfnamefont {M.}~\bibnamefont {Sasaki}},\ }\href {\doibase 10.1088/1475-7516/2018/05/042} {\bibfield  {journal} {\bibinfo  {journal} {JCAP}\ }\textbf {\bibinfo {volume} {05}},\ \bibinfo {pages} {042} (\bibinfo {year} {2018})}\BibitemShut {NoStop}%
\bibitem [{\citenamefont {Chen}\ \emph {et~al.}(2020)\citenamefont {Chen}, \citenamefont {Ma},\ and\ \citenamefont {Cai}}]{Chen:2020uhe}%
  \BibitemOpen
  \bibfield  {author} {\bibinfo {author} {\bibfnamefont {C.}~\bibnamefont {Chen}}, \bibinfo {author} {\bibfnamefont {X.-H.}\ \bibnamefont {Ma}}, \ and\ \bibinfo {author} {\bibfnamefont {Y.-F.}\ \bibnamefont {Cai}},\ }\href {\doibase 10.1103/PhysRevD.102.063526} {\bibfield  {journal} {\bibinfo  {journal} {Phys. Rev. D}\ }\textbf {\bibinfo {volume} {102}},\ \bibinfo {pages} {063526} (\bibinfo {year} {2020})}\BibitemShut {NoStop}%
\bibitem [{\citenamefont {Carr}\ \emph {et~al.}(2021)\citenamefont {Carr}, \citenamefont {Kohri}, \citenamefont {Sendouda},\ and\ \citenamefont {Yokoyama}}]{carr_constraints_2021}%
  \BibitemOpen
  \bibfield  {author} {\bibinfo {author} {\bibfnamefont {B.}~\bibnamefont {Carr}}, \bibinfo {author} {\bibfnamefont {K.}~\bibnamefont {Kohri}}, \bibinfo {author} {\bibfnamefont {Y.}~\bibnamefont {Sendouda}}, \ and\ \bibinfo {author} {\bibfnamefont {J.}~\bibnamefont {Yokoyama}},\ }\href {\doibase 10.1088/1361-6633/ac1e31} {\bibfield  {journal} {\bibinfo  {journal} {Rept. Prog. Phys.}\ }\textbf {\bibinfo {volume} {84}},\ \bibinfo {pages} {116902} (\bibinfo {year} {2021})}\BibitemShut {NoStop}%
\bibitem [{\citenamefont {Cai}\ \emph {et~al.}(2018)\citenamefont {Cai}, \citenamefont {Tong}, \citenamefont {Wang},\ and\ \citenamefont {Yan}}]{Cai:2018tuh}%
  \BibitemOpen
  \bibfield  {author} {\bibinfo {author} {\bibfnamefont {Y.-F.}\ \bibnamefont {Cai}}, \bibinfo {author} {\bibfnamefont {X.}~\bibnamefont {Tong}}, \bibinfo {author} {\bibfnamefont {D.-G.}\ \bibnamefont {Wang}}, \ and\ \bibinfo {author} {\bibfnamefont {S.-F.}\ \bibnamefont {Yan}},\ }\href {\doibase 10.1103/PhysRevLett.121.081306} {\bibfield  {journal} {\bibinfo  {journal} {Phys. Rev. Lett.}\ }\textbf {\bibinfo {volume} {121}},\ \bibinfo {pages} {081306} (\bibinfo {year} {2018})}\BibitemShut {NoStop}%
\bibitem [{\citenamefont {Cai}\ \emph {et~al.}(2021{\natexlab{b}})\citenamefont {Cai}, \citenamefont {Lin}, \citenamefont {Wang},\ and\ \citenamefont {Yan}}]{Cai:2020ovp}%
  \BibitemOpen
  \bibfield  {author} {\bibinfo {author} {\bibfnamefont {Y.-F.}\ \bibnamefont {Cai}}, \bibinfo {author} {\bibfnamefont {C.}~\bibnamefont {Lin}}, \bibinfo {author} {\bibfnamefont {B.}~\bibnamefont {Wang}}, \ and\ \bibinfo {author} {\bibfnamefont {S.-F.}\ \bibnamefont {Yan}},\ }\href {\doibase 10.1103/PhysRevLett.126.071303} {\bibfield  {journal} {\bibinfo  {journal} {Phys. Rev. Lett.}\ }\textbf {\bibinfo {volume} {126}},\ \bibinfo {pages} {071303} (\bibinfo {year} {2021}{\natexlab{b}})}\BibitemShut {NoStop}%
\bibitem [{\citenamefont {Cai}\ \emph {et~al.}(2021{\natexlab{c}})\citenamefont {Cai}, \citenamefont {Chen},\ and\ \citenamefont {Fu}}]{Cai:2021wzd}%
  \BibitemOpen
  \bibfield  {author} {\bibinfo {author} {\bibfnamefont {R.-G.}\ \bibnamefont {Cai}}, \bibinfo {author} {\bibfnamefont {C.}~\bibnamefont {Chen}}, \ and\ \bibinfo {author} {\bibfnamefont {C.}~\bibnamefont {Fu}},\ }\href {\doibase 10.1103/PhysRevD.104.083537} {\bibfield  {journal} {\bibinfo  {journal} {Phys. Rev. D}\ }\textbf {\bibinfo {volume} {104}},\ \bibinfo {pages} {083537} (\bibinfo {year} {2021}{\natexlab{c}})}\BibitemShut {NoStop}%
\bibitem [{\citenamefont {Cai}\ \emph {et~al.}(2024)\citenamefont {Cai}, \citenamefont {Tang}, \citenamefont {Mo}, \citenamefont {Yan}, \citenamefont {Chen}, \citenamefont {Ma}, \citenamefont {Wang}, \citenamefont {Luo}, \citenamefont {Easson},\ and\ \citenamefont {Marciano}}]{Cai:2023ptf}%
  \BibitemOpen
  \bibfield  {author} {\bibinfo {author} {\bibfnamefont {Y.-F.}\ \bibnamefont {Cai}}, \bibinfo {author} {\bibfnamefont {C.}~\bibnamefont {Tang}}, \bibinfo {author} {\bibfnamefont {G.}~\bibnamefont {Mo}}, \bibinfo {author} {\bibfnamefont {S.-F.}\ \bibnamefont {Yan}}, \bibinfo {author} {\bibfnamefont {C.}~\bibnamefont {Chen}}, \bibinfo {author} {\bibfnamefont {X.-H.}\ \bibnamefont {Ma}}, \bibinfo {author} {\bibfnamefont {B.}~\bibnamefont {Wang}}, \bibinfo {author} {\bibfnamefont {W.}~\bibnamefont {Luo}}, \bibinfo {author} {\bibfnamefont {D.~A.}\ \bibnamefont {Easson}}, \ and\ \bibinfo {author} {\bibfnamefont {A.}~\bibnamefont {Marciano}},\ }\href {\doibase 10.1007/s11433-023-2314-1} {\bibfield  {journal} {\bibinfo  {journal} {Sci. China Phys. Mech. Astron.}\ }\textbf {\bibinfo {volume} {67}},\ \bibinfo {pages} {259512} (\bibinfo {year} {2024})}\BibitemShut {NoStop}%
\bibitem [{\citenamefont {Kawasaki}\ \emph {et~al.}(2012)\citenamefont {Kawasaki}, \citenamefont {Kusenko},\ and\ \citenamefont {Yanagida}}]{kawasaki_primordial_2012}%
  \BibitemOpen
  \bibfield  {author} {\bibinfo {author} {\bibfnamefont {M.}~\bibnamefont {Kawasaki}}, \bibinfo {author} {\bibfnamefont {A.}~\bibnamefont {Kusenko}}, \ and\ \bibinfo {author} {\bibfnamefont {T.~T.}\ \bibnamefont {Yanagida}},\ }\href {\doibase 10.1016/j.physletb.2012.03.056} {\bibfield  {journal} {\bibinfo  {journal} {Phys. Lett. B}\ }\textbf {\bibinfo {volume} {711}},\ \bibinfo {pages} {1} (\bibinfo {year} {2012})}\BibitemShut {NoStop}%
\bibitem [{\citenamefont {Hooper}\ \emph {et~al.}(2024)\citenamefont {Hooper}, \citenamefont {Ireland}, \citenamefont {Krnjaic},\ and\ \citenamefont {Stebbins}}]{Hooper_2024}%
  \BibitemOpen
  \bibfield  {author} {\bibinfo {author} {\bibfnamefont {D.}~\bibnamefont {Hooper}}, \bibinfo {author} {\bibfnamefont {A.}~\bibnamefont {Ireland}}, \bibinfo {author} {\bibfnamefont {G.}~\bibnamefont {Krnjaic}}, \ and\ \bibinfo {author} {\bibfnamefont {A.}~\bibnamefont {Stebbins}},\ }\href {\doibase 10.1088/1475-7516/2024/04/021} {\bibfield  {journal} {\bibinfo  {journal} {JCAP}\ }\textbf {\bibinfo {volume} {2024}},\ \bibinfo {pages} {021} (\bibinfo {year} {2024})}\BibitemShut {NoStop}%
\bibitem [{\citenamefont {Labbé}\ \emph {et~al.}(2023)\citenamefont {Labbé}, \citenamefont {van Dokkum}, \citenamefont {Nelson}, \citenamefont {Bezanson}, \citenamefont {Suess}, \citenamefont {Leja}, \citenamefont {Brammer}, \citenamefont {Whitaker}, \citenamefont {Mathews}, \citenamefont {Stefanon},\ and\ \citenamefont {Wang}}]{labbe_population_2023}%
  \BibitemOpen
  \bibfield  {author} {\bibinfo {author} {\bibfnamefont {I.}~\bibnamefont {Labbé}}, \bibinfo {author} {\bibfnamefont {P.}~\bibnamefont {van Dokkum}}, \bibinfo {author} {\bibfnamefont {E.}~\bibnamefont {Nelson}}, \bibinfo {author} {\bibfnamefont {R.}~\bibnamefont {Bezanson}}, \bibinfo {author} {\bibfnamefont {K.~A.}\ \bibnamefont {Suess}}, \bibinfo {author} {\bibfnamefont {J.}~\bibnamefont {Leja}}, \bibinfo {author} {\bibfnamefont {G.}~\bibnamefont {Brammer}}, \bibinfo {author} {\bibfnamefont {K.}~\bibnamefont {Whitaker}}, \bibinfo {author} {\bibfnamefont {E.}~\bibnamefont {Mathews}}, \bibinfo {author} {\bibfnamefont {M.}~\bibnamefont {Stefanon}}, \ and\ \bibinfo {author} {\bibfnamefont {B.}~\bibnamefont {Wang}},\ }\href {\doibase 10.1038/s41586-023-05786-2} {\bibfield  {journal} {\bibinfo  {journal} {Nature}\ }\textbf {\bibinfo {volume} {616}},\ \bibinfo {pages} {266} (\bibinfo {year} {2023})}\BibitemShut {NoStop}%
\bibitem [{\citenamefont {Endsley}\ \emph {et~al.}(2023)\citenamefont {Endsley}, \citenamefont {Stark}, \citenamefont {Lyu}, \citenamefont {Wang}, \citenamefont {Yang}, \citenamefont {Fan}, \citenamefont {Smit}, \citenamefont {Bouwens}, \citenamefont {Hainline},\ and\ \citenamefont {Schouws}}]{10.1093/mnras/stad266}%
  \BibitemOpen
  \bibfield  {author} {\bibinfo {author} {\bibfnamefont {R.}~\bibnamefont {Endsley}}, \bibinfo {author} {\bibfnamefont {D.~P.}\ \bibnamefont {Stark}}, \bibinfo {author} {\bibfnamefont {J.}~\bibnamefont {Lyu}}, \bibinfo {author} {\bibfnamefont {F.}~\bibnamefont {Wang}}, \bibinfo {author} {\bibfnamefont {J.}~\bibnamefont {Yang}}, \bibinfo {author} {\bibfnamefont {X.}~\bibnamefont {Fan}}, \bibinfo {author} {\bibfnamefont {R.}~\bibnamefont {Smit}}, \bibinfo {author} {\bibfnamefont {R.}~\bibnamefont {Bouwens}}, \bibinfo {author} {\bibfnamefont {K.}~\bibnamefont {Hainline}}, \ and\ \bibinfo {author} {\bibfnamefont {S.}~\bibnamefont {Schouws}},\ }\href {\doibase 10.1093/mnras/stad266} {\bibfield  {journal} {\bibinfo  {journal} {Mon. Not. R. Astron. Soc.}\ }\textbf {\bibinfo {volume} {520}},\ \bibinfo {pages} {4609} (\bibinfo {year} {2023})}\BibitemShut {NoStop}%
\bibitem [{\citenamefont {Liu}\ \emph {et~al.}(2024)\citenamefont {Liu}, \citenamefont {Chen},\ and\ \citenamefont {Huang}}]{Liu:2023ymk}%
  \BibitemOpen
  \bibfield  {author} {\bibinfo {author} {\bibfnamefont {L.}~\bibnamefont {Liu}}, \bibinfo {author} {\bibfnamefont {Z.-C.}\ \bibnamefont {Chen}}, \ and\ \bibinfo {author} {\bibfnamefont {Q.-G.}\ \bibnamefont {Huang}},\ }\href {\doibase 10.1103/PhysRevD.109.L061301} {\bibfield  {journal} {\bibinfo  {journal} {Phys. Rev. D}\ }\textbf {\bibinfo {volume} {109}},\ \bibinfo {pages} {L061301} (\bibinfo {year} {2024})}\BibitemShut {NoStop}%
\bibitem [{\citenamefont {Carr}\ and\ \citenamefont {Lidsey}(1993)}]{Carr:1993aq}%
  \BibitemOpen
  \bibfield  {author} {\bibinfo {author} {\bibfnamefont {B.~J.}\ \bibnamefont {Carr}}\ and\ \bibinfo {author} {\bibfnamefont {J.~E.}\ \bibnamefont {Lidsey}},\ }\href {\doibase 10.1103/PhysRevD.48.543} {\bibfield  {journal} {\bibinfo  {journal} {Phys. Rev. D}\ }\textbf {\bibinfo {volume} {48}},\ \bibinfo {pages} {543} (\bibinfo {year} {1993})}\BibitemShut {NoStop}%
\bibitem [{\citenamefont {Khatri}\ and\ \citenamefont {Sunyaev}(2012{\natexlab{a}})}]{khatri_creation_2012}%
  \BibitemOpen
  \bibfield  {author} {\bibinfo {author} {\bibfnamefont {R.}~\bibnamefont {Khatri}}\ and\ \bibinfo {author} {\bibfnamefont {R.~A.}\ \bibnamefont {Sunyaev}},\ }\href {\doibase 10.1088/1475-7516/2012/06/038} {\bibfield  {journal} {\bibinfo  {journal} {JCAP}\ }\textbf {\bibinfo {volume} {2012}},\ \bibinfo {pages} {038} (\bibinfo {year} {2012}{\natexlab{a}})}\BibitemShut {NoStop}%
\bibitem [{\citenamefont {Khatri}\ and\ \citenamefont {Sunyaev}(2012{\natexlab{b}})}]{khatri_beyond_2012}%
  \BibitemOpen
  \bibfield  {author} {\bibinfo {author} {\bibfnamefont {R.}~\bibnamefont {Khatri}}\ and\ \bibinfo {author} {\bibfnamefont {R.~A.}\ \bibnamefont {Sunyaev}},\ }\href {\doibase 10.1088/1475-7516/2012/09/016} {\bibfield  {journal} {\bibinfo  {journal} {JCAP}\ }\textbf {\bibinfo {volume} {2012}},\ \bibinfo {pages} {016} (\bibinfo {year} {2012}{\natexlab{b}})}\BibitemShut {NoStop}%
\bibitem [{\citenamefont {Sunyaev}\ and\ \citenamefont {Zeldovich}(1970)}]{sunyaev_small_1970}%
  \BibitemOpen
  \bibfield  {author} {\bibinfo {author} {\bibfnamefont {R.~A.}\ \bibnamefont {Sunyaev}}\ and\ \bibinfo {author} {\bibfnamefont {Y.~B.}\ \bibnamefont {Zeldovich}},\ }\href {\doibase 10.1007/BF00649577} {\bibfield  {journal} {\bibinfo  {journal} {Astrophys. Space Sci.}\ }\textbf {\bibinfo {volume} {9}},\ \bibinfo {pages} {368} (\bibinfo {year} {1970})}\BibitemShut {NoStop}%
\bibitem [{\citenamefont {Chluba}\ \emph {et~al.}(2012{\natexlab{a}})\citenamefont {Chluba}, \citenamefont {Erickcek},\ and\ \citenamefont {Ben-Dayan}}]{chluba_probing_2012}%
  \BibitemOpen
  \bibfield  {author} {\bibinfo {author} {\bibfnamefont {J.}~\bibnamefont {Chluba}}, \bibinfo {author} {\bibfnamefont {A.~L.}\ \bibnamefont {Erickcek}}, \ and\ \bibinfo {author} {\bibfnamefont {I.}~\bibnamefont {Ben-Dayan}},\ }\href {\doibase 10.1088/0004-637X/758/2/76} {\bibfield  {journal} {\bibinfo  {journal} {Astrophys. J.}\ }\textbf {\bibinfo {volume} {758}},\ \bibinfo {pages} {76} (\bibinfo {year} {2012}{\natexlab{a}})}\BibitemShut {NoStop}%
\bibitem [{\citenamefont {Lifshitz}(1946)}]{Lifshitz:1945du}%
  \BibitemOpen
  \bibfield  {author} {\bibinfo {author} {\bibfnamefont {E.}~\bibnamefont {Lifshitz}},\ }\href {\doibase 10.1007/s10714-016-2165-8} {\bibfield  {journal} {\bibinfo  {journal} {J. Phys. (USSR)}\ }\textbf {\bibinfo {volume} {10}},\ \bibinfo {pages} {116} (\bibinfo {year} {1946})}\BibitemShut {NoStop}%
\bibitem [{\citenamefont {Lifshitz}\ and\ \citenamefont {Khalatnikov}(1963)}]{Lifshitz:1963ps}%
  \BibitemOpen
  \bibfield  {author} {\bibinfo {author} {\bibfnamefont {E.~M.}\ \bibnamefont {Lifshitz}}\ and\ \bibinfo {author} {\bibfnamefont {I.~M.}\ \bibnamefont {Khalatnikov}},\ }\href {\doibase 10.1080/00018736300101283} {\bibfield  {journal} {\bibinfo  {journal} {Adv. Phys.}\ }\textbf {\bibinfo {volume} {12}},\ \bibinfo {pages} {185} (\bibinfo {year} {1963})}\BibitemShut {NoStop}%
\bibitem [{\citenamefont {Silk}(1968)}]{silk_cosmic_1968}%
  \BibitemOpen
  \bibfield  {author} {\bibinfo {author} {\bibfnamefont {J.}~\bibnamefont {Silk}},\ }\href {\doibase 10.1086/149449} {\bibfield  {journal} {\bibinfo  {journal} {Astrophys. J.}\ }\textbf {\bibinfo {volume} {151}},\ \bibinfo {pages} {459} (\bibinfo {year} {1968})}\BibitemShut {NoStop}%
\bibitem [{\citenamefont {Fixsen}\ \emph {et~al.}(1996)\citenamefont {Fixsen}, \citenamefont {Cheng}, \citenamefont {Gales}, \citenamefont {Mather}, \citenamefont {Shafer},\ and\ \citenamefont {Wright}}]{fixsen_cosmic_1996}%
  \BibitemOpen
  \bibfield  {author} {\bibinfo {author} {\bibfnamefont {D.~J.}\ \bibnamefont {Fixsen}}, \bibinfo {author} {\bibfnamefont {E.~S.}\ \bibnamefont {Cheng}}, \bibinfo {author} {\bibfnamefont {J.~M.}\ \bibnamefont {Gales}}, \bibinfo {author} {\bibfnamefont {J.~C.}\ \bibnamefont {Mather}}, \bibinfo {author} {\bibfnamefont {R.~A.}\ \bibnamefont {Shafer}}, \ and\ \bibinfo {author} {\bibfnamefont {E.~L.}\ \bibnamefont {Wright}},\ }\href {\doibase 10.1086/178173} {\bibfield  {journal} {\bibinfo  {journal} {Astrophys. J.}\ }\textbf {\bibinfo {volume} {473}},\ \bibinfo {pages} {576} (\bibinfo {year} {1996})}\BibitemShut {NoStop}%
\bibitem [{\citenamefont {Fixsen}\ \emph {et~al.}(1998)\citenamefont {Fixsen}, \citenamefont {Dwek}, \citenamefont {Mather}, \citenamefont {Bennett},\ and\ \citenamefont {Shafer}}]{fixsen_spectrum_1998}%
  \BibitemOpen
  \bibfield  {author} {\bibinfo {author} {\bibfnamefont {D.}~\bibnamefont {Fixsen}}, \bibinfo {author} {\bibfnamefont {E.}~\bibnamefont {Dwek}}, \bibinfo {author} {\bibfnamefont {J.}~\bibnamefont {Mather}}, \bibinfo {author} {\bibfnamefont {C.}~\bibnamefont {Bennett}}, \ and\ \bibinfo {author} {\bibfnamefont {R.}~\bibnamefont {Shafer}},\ }\href {\doibase 10.1086/306383} {\bibfield  {journal} {\bibinfo  {journal} {Astrophys. J.}\ }\textbf {\bibinfo {volume} {508}},\ \bibinfo {pages} {123} (\bibinfo {year} {1998})}\BibitemShut {NoStop}%
\bibitem [{\citenamefont {Kohri}\ \emph {et~al.}(2014)\citenamefont {Kohri}, \citenamefont {Nakama},\ and\ \citenamefont {Suyama}}]{kohri_testing_2014}%
  \BibitemOpen
  \bibfield  {author} {\bibinfo {author} {\bibfnamefont {K.}~\bibnamefont {Kohri}}, \bibinfo {author} {\bibfnamefont {T.}~\bibnamefont {Nakama}}, \ and\ \bibinfo {author} {\bibfnamefont {T.}~\bibnamefont {Suyama}},\ }\href {\doibase 10.1103/PhysRevD.90.083514} {\bibfield  {journal} {\bibinfo  {journal} {Phys. Rev. D}\ }\textbf {\bibinfo {volume} {90}},\ \bibinfo {pages} {083514} (\bibinfo {year} {2014})}\BibitemShut {NoStop}%
\bibitem [{\citenamefont {Nakama}\ \emph {et~al.}(2016)\citenamefont {Nakama}, \citenamefont {Suyama},\ and\ \citenamefont {Yokoyama}}]{PhysRevD.94.103522}%
  \BibitemOpen
  \bibfield  {author} {\bibinfo {author} {\bibfnamefont {T.}~\bibnamefont {Nakama}}, \bibinfo {author} {\bibfnamefont {T.}~\bibnamefont {Suyama}}, \ and\ \bibinfo {author} {\bibfnamefont {J.}~\bibnamefont {Yokoyama}},\ }\href {\doibase 10.1103/PhysRevD.94.103522} {\bibfield  {journal} {\bibinfo  {journal} {Phys. Rev. D}\ }\textbf {\bibinfo {volume} {94}},\ \bibinfo {pages} {103522} (\bibinfo {year} {2016})}\BibitemShut {NoStop}%
\bibitem [{\citenamefont {Kawasaki}\ and\ \citenamefont {Murai}(2019)}]{PhysRevD.100.103521}%
  \BibitemOpen
  \bibfield  {author} {\bibinfo {author} {\bibfnamefont {M.}~\bibnamefont {Kawasaki}}\ and\ \bibinfo {author} {\bibfnamefont {K.}~\bibnamefont {Murai}},\ }\href {\doibase 10.1103/PhysRevD.100.103521} {\bibfield  {journal} {\bibinfo  {journal} {Phys. Rev. D}\ }\textbf {\bibinfo {volume} {100}},\ \bibinfo {pages} {103521} (\bibinfo {year} {2019})}\BibitemShut {NoStop}%
\bibitem [{\citenamefont {Abitbol}\ \emph {et~al.}(2017)\citenamefont {Abitbol}, \citenamefont {Chluba}, \citenamefont {Hill},\ and\ \citenamefont {Johnson}}]{abitbol_prospects_2017}%
  \BibitemOpen
  \bibfield  {author} {\bibinfo {author} {\bibfnamefont {M.~H.}\ \bibnamefont {Abitbol}}, \bibinfo {author} {\bibfnamefont {J.}~\bibnamefont {Chluba}}, \bibinfo {author} {\bibfnamefont {J.~C.}\ \bibnamefont {Hill}}, \ and\ \bibinfo {author} {\bibfnamefont {B.~R.}\ \bibnamefont {Johnson}},\ }\href {\doibase 10.1093/mnras/stx1653} {\bibfield  {journal} {\bibinfo  {journal} {Mon. Not. R. Astron. Soc.}\ }\textbf {\bibinfo {volume} {471}},\ \bibinfo {pages} {1126} (\bibinfo {year} {2017})}\BibitemShut {NoStop}%
\bibitem [{\citenamefont {García-Bellido}\ \emph {et~al.}(2017)\citenamefont {García-Bellido}, \citenamefont {Peloso},\ and\ \citenamefont {Unal}}]{García-Bellido_2017}%
  \BibitemOpen
  \bibfield  {author} {\bibinfo {author} {\bibfnamefont {J.}~\bibnamefont {García-Bellido}}, \bibinfo {author} {\bibfnamefont {M.}~\bibnamefont {Peloso}}, \ and\ \bibinfo {author} {\bibfnamefont {C.}~\bibnamefont {Unal}},\ }\href {\doibase 10.1088/1475-7516/2017/09/013} {\bibfield  {journal} {\bibinfo  {journal} {JCAP}\ }\textbf {\bibinfo {volume} {2017}},\ \bibinfo {pages} {013} (\bibinfo {year} {2017})}\BibitemShut {NoStop}%
\bibitem [{\citenamefont {Deng}(2021)}]{Deng_2021}%
  \BibitemOpen
  \bibfield  {author} {\bibinfo {author} {\bibfnamefont {H.}~\bibnamefont {Deng}},\ }\href {\doibase 10.1088/1475-7516/2021/11/054} {\bibfield  {journal} {\bibinfo  {journal} {JCAP}\ }\textbf {\bibinfo {volume} {2021}},\ \bibinfo {pages} {054} (\bibinfo {year} {2021})}\BibitemShut {NoStop}%
\bibitem [{\citenamefont {Acharya}\ and\ \citenamefont {Khatri}(2020)}]{Acharya_2020}%
  \BibitemOpen
  \bibfield  {author} {\bibinfo {author} {\bibfnamefont {S.~K.}\ \bibnamefont {Acharya}}\ and\ \bibinfo {author} {\bibfnamefont {R.}~\bibnamefont {Khatri}},\ }\href {\doibase 10.1088/1475-7516/2020/02/010} {\bibfield  {journal} {\bibinfo  {journal} {JCAP}\ }\textbf {\bibinfo {volume} {2020}},\ \bibinfo {pages} {010} (\bibinfo {year} {2020})}\BibitemShut {NoStop}%
\bibitem [{\citenamefont {Yang}(2022)}]{yang_constraints_2022}%
  \BibitemOpen
  \bibfield  {author} {\bibinfo {author} {\bibfnamefont {Y.}~\bibnamefont {Yang}},\ }\href {\doibase 10.1103/PhysRevD.106.043516} {\bibfield  {journal} {\bibinfo  {journal} {Phys. Rev. D}\ }\textbf {\bibinfo {volume} {106}},\ \bibinfo {pages} {043516} (\bibinfo {year} {2022})}\BibitemShut {NoStop}%
\bibitem [{\citenamefont {Zegeye}\ \emph {et~al.}(2022)\citenamefont {Zegeye}, \citenamefont {Inomata},\ and\ \citenamefont {Hu}}]{PhysRevD.105.103535}%
  \BibitemOpen
  \bibfield  {author} {\bibinfo {author} {\bibfnamefont {D.}~\bibnamefont {Zegeye}}, \bibinfo {author} {\bibfnamefont {K.}~\bibnamefont {Inomata}}, \ and\ \bibinfo {author} {\bibfnamefont {W.}~\bibnamefont {Hu}},\ }\href {\doibase 10.1103/PhysRevD.105.103535} {\bibfield  {journal} {\bibinfo  {journal} {Phys. Rev. D}\ }\textbf {\bibinfo {volume} {105}},\ \bibinfo {pages} {103535} (\bibinfo {year} {2022})}\BibitemShut {NoStop}%
\bibitem [{\citenamefont {De~Luca}\ and\ \citenamefont {Riotto}(2022)}]{de_luca_note_2022}%
  \BibitemOpen
  \bibfield  {author} {\bibinfo {author} {\bibfnamefont {V.}~\bibnamefont {De~Luca}}\ and\ \bibinfo {author} {\bibfnamefont {A.}~\bibnamefont {Riotto}},\ }\href {\doibase 10.1016/j.physletb.2022.137035} {\bibfield  {journal} {\bibinfo  {journal} {Phys. Lett. B}\ }\textbf {\bibinfo {volume} {828}},\ \bibinfo {pages} {137035} (\bibinfo {year} {2022})}\BibitemShut {NoStop}%
\bibitem [{\citenamefont {Escriv\`a}\ \emph {et~al.}(2020)\citenamefont {Escriv\`a}, \citenamefont {Germani},\ and\ \citenamefont {Sheth}}]{PhysRevD.101.044022}%
  \BibitemOpen
  \bibfield  {author} {\bibinfo {author} {\bibfnamefont {A.}~\bibnamefont {Escriv\`a}}, \bibinfo {author} {\bibfnamefont {C.}~\bibnamefont {Germani}}, \ and\ \bibinfo {author} {\bibfnamefont {R.~K.}\ \bibnamefont {Sheth}},\ }\href {\doibase 10.1103/PhysRevD.101.044022} {\bibfield  {journal} {\bibinfo  {journal} {Phys. Rev. D}\ }\textbf {\bibinfo {volume} {101}},\ \bibinfo {pages} {044022} (\bibinfo {year} {2020})}\BibitemShut {NoStop}%
\bibitem [{\citenamefont {Escrivà}\ \emph {et~al.}(2021)\citenamefont {Escrivà}, \citenamefont {Germani},\ and\ \citenamefont {Sheth}}]{Escrivà_2021}%
  \BibitemOpen
  \bibfield  {author} {\bibinfo {author} {\bibfnamefont {A.}~\bibnamefont {Escrivà}}, \bibinfo {author} {\bibfnamefont {C.}~\bibnamefont {Germani}}, \ and\ \bibinfo {author} {\bibfnamefont {R.~K.}\ \bibnamefont {Sheth}},\ }\href {\doibase 10.1088/1475-7516/2021/01/030} {\bibfield  {journal} {\bibinfo  {journal} {JCAP}\ }\textbf {\bibinfo {volume} {2021}},\ \bibinfo {pages} {030} (\bibinfo {year} {2021})}\BibitemShut {NoStop}%
\bibitem [{\citenamefont {Musco}\ \emph {et~al.}(2021)\citenamefont {Musco}, \citenamefont {De~Luca}, \citenamefont {Franciolini},\ and\ \citenamefont {Riotto}}]{PhysRevD.103.063538}%
  \BibitemOpen
  \bibfield  {author} {\bibinfo {author} {\bibfnamefont {I.}~\bibnamefont {Musco}}, \bibinfo {author} {\bibfnamefont {V.}~\bibnamefont {De~Luca}}, \bibinfo {author} {\bibfnamefont {G.}~\bibnamefont {Franciolini}}, \ and\ \bibinfo {author} {\bibfnamefont {A.}~\bibnamefont {Riotto}},\ }\href {\doibase 10.1103/PhysRevD.103.063538} {\bibfield  {journal} {\bibinfo  {journal} {Phys. Rev. D}\ }\textbf {\bibinfo {volume} {103}},\ \bibinfo {pages} {063538} (\bibinfo {year} {2021})}\BibitemShut {NoStop}%
\bibitem [{\citenamefont {Harada}\ \emph {et~al.}(2015)\citenamefont {Harada}, \citenamefont {Yoo}, \citenamefont {Nakama},\ and\ \citenamefont {Koga}}]{PhysRevD.91.084057}%
  \BibitemOpen
  \bibfield  {author} {\bibinfo {author} {\bibfnamefont {T.}~\bibnamefont {Harada}}, \bibinfo {author} {\bibfnamefont {C.-M.}\ \bibnamefont {Yoo}}, \bibinfo {author} {\bibfnamefont {T.}~\bibnamefont {Nakama}}, \ and\ \bibinfo {author} {\bibfnamefont {Y.}~\bibnamefont {Koga}},\ }\href {\doibase 10.1103/PhysRevD.91.084057} {\bibfield  {journal} {\bibinfo  {journal} {Phys. Rev. D}\ }\textbf {\bibinfo {volume} {91}},\ \bibinfo {pages} {084057} (\bibinfo {year} {2015})}\BibitemShut {NoStop}%
\bibitem [{\citenamefont {Germani}\ and\ \citenamefont {Musco}(2019)}]{PhysRevLett.122.141302}%
  \BibitemOpen
  \bibfield  {author} {\bibinfo {author} {\bibfnamefont {C.}~\bibnamefont {Germani}}\ and\ \bibinfo {author} {\bibfnamefont {I.}~\bibnamefont {Musco}},\ }\href {\doibase 10.1103/PhysRevLett.122.141302} {\bibfield  {journal} {\bibinfo  {journal} {Phys. Rev. Lett.}\ }\textbf {\bibinfo {volume} {122}},\ \bibinfo {pages} {141302} (\bibinfo {year} {2019})}\BibitemShut {NoStop}%
\bibitem [{\citenamefont {Young}\ \emph {et~al.}(2019)\citenamefont {Young}, \citenamefont {Musco},\ and\ \citenamefont {Byrnes}}]{Young_2019}%
  \BibitemOpen
  \bibfield  {author} {\bibinfo {author} {\bibfnamefont {S.}~\bibnamefont {Young}}, \bibinfo {author} {\bibfnamefont {I.}~\bibnamefont {Musco}}, \ and\ \bibinfo {author} {\bibfnamefont {C.~T.}\ \bibnamefont {Byrnes}},\ }\href {\doibase 10.1088/1475-7516/2019/11/012} {\bibfield  {journal} {\bibinfo  {journal} {JCAP}\ }\textbf {\bibinfo {volume} {2019}},\ \bibinfo {pages} {012} (\bibinfo {year} {2019})}\BibitemShut {NoStop}%
\bibitem [{\citenamefont {Musco}(2019)}]{musco_threshold_2019}%
  \BibitemOpen
  \bibfield  {author} {\bibinfo {author} {\bibfnamefont {I.}~\bibnamefont {Musco}},\ }\href {\doibase 10.1103/PhysRevD.100.123524} {\bibfield  {journal} {\bibinfo  {journal} {Phys. Rev. D}\ }\textbf {\bibinfo {volume} {100}},\ \bibinfo {pages} {123524} (\bibinfo {year} {2019})}\BibitemShut {NoStop}%
\bibitem [{\citenamefont {Nakama}\ \emph {et~al.}(2018)\citenamefont {Nakama}, \citenamefont {Carr},\ and\ \citenamefont {Silk}}]{nakama_limits_2018}%
  \BibitemOpen
  \bibfield  {author} {\bibinfo {author} {\bibfnamefont {T.}~\bibnamefont {Nakama}}, \bibinfo {author} {\bibfnamefont {B.}~\bibnamefont {Carr}}, \ and\ \bibinfo {author} {\bibfnamefont {J.}~\bibnamefont {Silk}},\ }\href {\doibase 10.1103/PhysRevD.97.043525} {\bibfield  {journal} {\bibinfo  {journal} {Phys. Rev. D}\ }\textbf {\bibinfo {volume} {97}},\ \bibinfo {pages} {043525} (\bibinfo {year} {2018})}\BibitemShut {NoStop}%
\bibitem [{\citenamefont {Press}\ and\ \citenamefont {Schechter}(1974)}]{Press:1973iz}%
  \BibitemOpen
  \bibfield  {author} {\bibinfo {author} {\bibfnamefont {W.~H.}\ \bibnamefont {Press}}\ and\ \bibinfo {author} {\bibfnamefont {P.}~\bibnamefont {Schechter}},\ }\href {\doibase 10.1086/152650} {\bibfield  {journal} {\bibinfo  {journal} {Astrophys. J.}\ }\textbf {\bibinfo {volume} {187}},\ \bibinfo {pages} {425} (\bibinfo {year} {1974})}\BibitemShut {NoStop}%
\bibitem [{\citenamefont {Green}\ \emph {et~al.}(2004)\citenamefont {Green}, \citenamefont {Liddle}, \citenamefont {Malik},\ and\ \citenamefont {Sasaki}}]{green_new_2004}%
  \BibitemOpen
  \bibfield  {author} {\bibinfo {author} {\bibfnamefont {A.~M.}\ \bibnamefont {Green}}, \bibinfo {author} {\bibfnamefont {A.~R.}\ \bibnamefont {Liddle}}, \bibinfo {author} {\bibfnamefont {K.~A.}\ \bibnamefont {Malik}}, \ and\ \bibinfo {author} {\bibfnamefont {M.}~\bibnamefont {Sasaki}},\ }\href {\doibase 10.1103/PhysRevD.70.041502} {\bibfield  {journal} {\bibinfo  {journal} {Phys. Rev. D}\ }\textbf {\bibinfo {volume} {70}},\ \bibinfo {pages} {041502} (\bibinfo {year} {2004})}\BibitemShut {NoStop}%
\bibitem [{\citenamefont {Young}\ \emph {et~al.}(2014)\citenamefont {Young}, \citenamefont {Byrnes},\ and\ \citenamefont {Sasaki}}]{young_calculating_2014}%
  \BibitemOpen
  \bibfield  {author} {\bibinfo {author} {\bibfnamefont {S.}~\bibnamefont {Young}}, \bibinfo {author} {\bibfnamefont {C.~T.}\ \bibnamefont {Byrnes}}, \ and\ \bibinfo {author} {\bibfnamefont {M.}~\bibnamefont {Sasaki}},\ }\href {\doibase 10.1088/1475-7516/2014/07/045} {\bibfield  {journal} {\bibinfo  {journal} {JCAP}\ }\textbf {\bibinfo {volume} {2014}},\ \bibinfo {pages} {045} (\bibinfo {year} {2014})}\BibitemShut {NoStop}%
\bibitem [{\citenamefont {Kitajima}\ \emph {et~al.}(2021)\citenamefont {Kitajima}, \citenamefont {Tada}, \citenamefont {Yokoyama},\ and\ \citenamefont {Yoo}}]{kitajima_primordial_2021}%
  \BibitemOpen
  \bibfield  {author} {\bibinfo {author} {\bibfnamefont {N.}~\bibnamefont {Kitajima}}, \bibinfo {author} {\bibfnamefont {Y.}~\bibnamefont {Tada}}, \bibinfo {author} {\bibfnamefont {S.}~\bibnamefont {Yokoyama}}, \ and\ \bibinfo {author} {\bibfnamefont {C.-M.}\ \bibnamefont {Yoo}},\ }\href {\doibase 10.1088/1475-7516/2021/10/053} {\bibfield  {journal} {\bibinfo  {journal} {JCAP}\ }\textbf {\bibinfo {volume} {2021}},\ \bibinfo {pages} {053} (\bibinfo {year} {2021})}\BibitemShut {NoStop}%
\bibitem [{\citenamefont {Sureda}\ \emph {et~al.}(2021)\citenamefont {Sureda}, \citenamefont {Magana}, \citenamefont {Araya},\ and\ \citenamefont {Padilla}}]{sureda_press-schechter_2021}%
  \BibitemOpen
  \bibfield  {author} {\bibinfo {author} {\bibfnamefont {J.}~\bibnamefont {Sureda}}, \bibinfo {author} {\bibfnamefont {J.}~\bibnamefont {Magana}}, \bibinfo {author} {\bibfnamefont {I.~J.}\ \bibnamefont {Araya}}, \ and\ \bibinfo {author} {\bibfnamefont {N.~D.}\ \bibnamefont {Padilla}},\ }\href {\doibase 10.1093/mnras/stab2450} {\bibfield  {journal} {\bibinfo  {journal} {Mon. Not. R. Astron. Soc.}\ }\textbf {\bibinfo {volume} {507}},\ \bibinfo {pages} {4804} (\bibinfo {year} {2021})}\BibitemShut {NoStop}%
\bibitem [{\citenamefont {Aghanim}\ \emph {et~al.}(2020)\citenamefont {Aghanim} \emph {et~al.}}]{Planck:2018vyg}%
  \BibitemOpen
  \bibfield  {author} {\bibinfo {author} {\bibfnamefont {N.}~\bibnamefont {Aghanim}} \emph {et~al.} (\bibinfo {collaboration} {Planck}),\ }\href {\doibase 10.1051/0004-6361/201833910} {\bibfield  {journal} {\bibinfo  {journal} {Astron. Astrophys.}\ }\textbf {\bibinfo {volume} {641}},\ \bibinfo {pages} {A6} (\bibinfo {year} {2020})}\BibitemShut {NoStop}%
\bibitem [{\citenamefont {Carr}\ \emph {et~al.}(2010)\citenamefont {Carr}, \citenamefont {Kohri}, \citenamefont {Sendouda},\ and\ \citenamefont {Yokoyama}}]{Carr:2009jm}%
  \BibitemOpen
  \bibfield  {author} {\bibinfo {author} {\bibfnamefont {B.~J.}\ \bibnamefont {Carr}}, \bibinfo {author} {\bibfnamefont {K.}~\bibnamefont {Kohri}}, \bibinfo {author} {\bibfnamefont {Y.}~\bibnamefont {Sendouda}}, \ and\ \bibinfo {author} {\bibfnamefont {J.}~\bibnamefont {Yokoyama}},\ }\href {\doibase 10.1103/PhysRevD.81.104019} {\bibfield  {journal} {\bibinfo  {journal} {Phys. Rev. D}\ }\textbf {\bibinfo {volume} {81}},\ \bibinfo {pages} {104019} (\bibinfo {year} {2010})}\BibitemShut {NoStop}%
\bibitem [{\citenamefont {Sasaki}\ \emph {et~al.}(2018)\citenamefont {Sasaki}, \citenamefont {Suyama}, \citenamefont {Tanaka},\ and\ \citenamefont {Yokoyama}}]{sasaki_primordial_2018}%
  \BibitemOpen
  \bibfield  {author} {\bibinfo {author} {\bibfnamefont {M.}~\bibnamefont {Sasaki}}, \bibinfo {author} {\bibfnamefont {T.}~\bibnamefont {Suyama}}, \bibinfo {author} {\bibfnamefont {T.}~\bibnamefont {Tanaka}}, \ and\ \bibinfo {author} {\bibfnamefont {S.}~\bibnamefont {Yokoyama}},\ }\href {\doibase 10.1088/1361-6382/aaa7b4} {\bibfield  {journal} {\bibinfo  {journal} {Classical and Quantum Gravity}\ }\textbf {\bibinfo {volume} {35}},\ \bibinfo {pages} {063001} (\bibinfo {year} {2018})}\BibitemShut {NoStop}%
\bibitem [{\citenamefont {Chluba}(2016)}]{chluba_which_2016}%
  \BibitemOpen
  \bibfield  {author} {\bibinfo {author} {\bibfnamefont {J.}~\bibnamefont {Chluba}},\ }\href {\doibase 10.1093/mnras/stw945} {\bibfield  {journal} {\bibinfo  {journal} {Mon. Not. R. Astron. Soc.}\ }\textbf {\bibinfo {volume} {460}},\ \bibinfo {pages} {227} (\bibinfo {year} {2016})}\BibitemShut {NoStop}%
\bibitem [{\citenamefont {Novikov}\ and\ \citenamefont {Mihalchenko}(2023)}]{novikov_separation_2023}%
  \BibitemOpen
  \bibfield  {author} {\bibinfo {author} {\bibfnamefont {D.~I.}\ \bibnamefont {Novikov}}\ and\ \bibinfo {author} {\bibfnamefont {A.~O.}\ \bibnamefont {Mihalchenko}},\ }\href {\doibase 10.1103/PhysRevD.107.063506} {\bibfield  {journal} {\bibinfo  {journal} {Phys. Rev. D}\ }\textbf {\bibinfo {volume} {107}},\ \bibinfo {pages} {063506} (\bibinfo {year} {2023})}\BibitemShut {NoStop}%
\bibitem [{\citenamefont {Hu}\ and\ \citenamefont {Silk}(1993)}]{hu_thermalization_1993}%
  \BibitemOpen
  \bibfield  {author} {\bibinfo {author} {\bibfnamefont {W.}~\bibnamefont {Hu}}\ and\ \bibinfo {author} {\bibfnamefont {J.}~\bibnamefont {Silk}},\ }\href {\doibase 10.1103/PhysRevD.48.485} {\bibfield  {journal} {\bibinfo  {journal} {Phys. Rev. D}\ }\textbf {\bibinfo {volume} {48}},\ \bibinfo {pages} {485} (\bibinfo {year} {1993})}\BibitemShut {NoStop}%
\bibitem [{\citenamefont {Chluba}\ \emph {et~al.}(2013)\citenamefont {Chluba}, \citenamefont {Switzer}, \citenamefont {Nelson},\ and\ \citenamefont {Nagai}}]{chluba_sunyaevzeldovich_2013}%
  \BibitemOpen
  \bibfield  {author} {\bibinfo {author} {\bibfnamefont {J.}~\bibnamefont {Chluba}}, \bibinfo {author} {\bibfnamefont {E.}~\bibnamefont {Switzer}}, \bibinfo {author} {\bibfnamefont {K.}~\bibnamefont {Nelson}}, \ and\ \bibinfo {author} {\bibfnamefont {D.}~\bibnamefont {Nagai}},\ }\href {\doibase 10.1093/mnras/stt110} {\bibfield  {journal} {\bibinfo  {journal} {Mon. Not. R. Astron. Soc.}\ }\textbf {\bibinfo {volume} {430}},\ \bibinfo {pages} {3054} (\bibinfo {year} {2013})}\BibitemShut {NoStop}%
\bibitem [{\citenamefont {Khatri}\ and\ \citenamefont {Sunyaev}(2015)}]{khatri_limits_2015}%
  \BibitemOpen
  \bibfield  {author} {\bibinfo {author} {\bibfnamefont {R.}~\bibnamefont {Khatri}}\ and\ \bibinfo {author} {\bibfnamefont {R.}~\bibnamefont {Sunyaev}},\ }\href {\doibase 10.1088/1475-7516/2015/08/013} {\bibfield  {journal} {\bibinfo  {journal} {JCAP}\ }\textbf {\bibinfo {volume} {2015}},\ \bibinfo {pages} {013} (\bibinfo {year} {2015})}\BibitemShut {NoStop}%
\bibitem [{pla(2016)}]{planck_collaboration_planck_2016}%
  \BibitemOpen
  \href {\doibase 10.1051/0004-6361/201525826} {\bibfield  {journal} {\bibinfo  {journal} {A\&A}\ }\textbf {\bibinfo {volume} {594}},\ \bibinfo {pages} {A22} (\bibinfo {year} {2016})}\BibitemShut {NoStop}%
\bibitem [{\citenamefont {Erler}\ \emph {et~al.}(2018)\citenamefont {Erler}, \citenamefont {Basu}, \citenamefont {Chluba},\ and\ \citenamefont {Bertoldi}}]{erler_plancks_2018}%
  \BibitemOpen
  \bibfield  {author} {\bibinfo {author} {\bibfnamefont {J.}~\bibnamefont {Erler}}, \bibinfo {author} {\bibfnamefont {K.}~\bibnamefont {Basu}}, \bibinfo {author} {\bibfnamefont {J.}~\bibnamefont {Chluba}}, \ and\ \bibinfo {author} {\bibfnamefont {F.}~\bibnamefont {Bertoldi}},\ }\href {\doibase 10.1093/mnras/sty327} {\bibfield  {journal} {\bibinfo  {journal} {Mon. Not. R. Astron. Soc.}\ }\textbf {\bibinfo {volume} {476}},\ \bibinfo {pages} {3360} (\bibinfo {year} {2018})}\BibitemShut {NoStop}%
\bibitem [{\citenamefont {Bianchini}\ and\ \citenamefont {Fabbian}(2022)}]{bianchini_cmb_2022}%
  \BibitemOpen
  \bibfield  {author} {\bibinfo {author} {\bibfnamefont {F.}~\bibnamefont {Bianchini}}\ and\ \bibinfo {author} {\bibfnamefont {G.}~\bibnamefont {Fabbian}},\ }\href {\doibase 10.1103/PhysRevD.106.063527} {\bibfield  {journal} {\bibinfo  {journal} {Phys. Rev. D}\ }\textbf {\bibinfo {volume} {106}},\ \bibinfo {pages} {063527} (\bibinfo {year} {2022})}\BibitemShut {NoStop}%
\bibitem [{\citenamefont {Hill}\ \emph {et~al.}(2015)\citenamefont {Hill}, \citenamefont {Battaglia}, \citenamefont {Chluba}, \citenamefont {Ferraro}, \citenamefont {Schaan},\ and\ \citenamefont {Spergel}}]{hill_taking_2015}%
  \BibitemOpen
  \bibfield  {author} {\bibinfo {author} {\bibfnamefont {J.~C.}\ \bibnamefont {Hill}}, \bibinfo {author} {\bibfnamefont {N.}~\bibnamefont {Battaglia}}, \bibinfo {author} {\bibfnamefont {J.}~\bibnamefont {Chluba}}, \bibinfo {author} {\bibfnamefont {S.}~\bibnamefont {Ferraro}}, \bibinfo {author} {\bibfnamefont {E.}~\bibnamefont {Schaan}}, \ and\ \bibinfo {author} {\bibfnamefont {D.~N.}\ \bibnamefont {Spergel}},\ }\href {\doibase 10.1103/PhysRevLett.115.261301} {\bibfield  {journal} {\bibinfo  {journal} {Phys. Rev. Lett.}\ }\textbf {\bibinfo {volume} {115}},\ \bibinfo {pages} {261301} (\bibinfo {year} {2015})}\BibitemShut {NoStop}%
\bibitem [{\citenamefont {Chluba}\ and\ \citenamefont {Sunyaev}(2012)}]{chluba_evolution_2012}%
  \BibitemOpen
  \bibfield  {author} {\bibinfo {author} {\bibfnamefont {J.}~\bibnamefont {Chluba}}\ and\ \bibinfo {author} {\bibfnamefont {R.~A.}\ \bibnamefont {Sunyaev}},\ }\href {\doibase 10.1111/j.1365-2966.2011.19786.x} {\bibfield  {journal} {\bibinfo  {journal} {Mon. Not. R. Astron. Soc.}\ }\textbf {\bibinfo {volume} {419}},\ \bibinfo {pages} {1294} (\bibinfo {year} {2012})}\BibitemShut {NoStop}%
\bibitem [{\citenamefont {Chluba}\ \emph {et~al.}(2012{\natexlab{b}})\citenamefont {Chluba}, \citenamefont {Khatri},\ and\ \citenamefont {Sunyaev}}]{chluba_cmb_2012}%
  \BibitemOpen
  \bibfield  {author} {\bibinfo {author} {\bibfnamefont {J.}~\bibnamefont {Chluba}}, \bibinfo {author} {\bibfnamefont {R.}~\bibnamefont {Khatri}}, \ and\ \bibinfo {author} {\bibfnamefont {R.~A.}\ \bibnamefont {Sunyaev}},\ }\href {\doibase 10.1111/j.1365-2966.2012.21474.x} {\bibfield  {journal} {\bibinfo  {journal} {Mon. Not. R. Astron. Soc.}\ }\textbf {\bibinfo {volume} {425}},\ \bibinfo {pages} {1129} (\bibinfo {year} {2012}{\natexlab{b}})}\BibitemShut {NoStop}%
\bibitem [{\citenamefont {Chluba}(2013{\natexlab{a}})}]{chluba_greens_2013}%
  \BibitemOpen
  \bibfield  {author} {\bibinfo {author} {\bibfnamefont {J.}~\bibnamefont {Chluba}},\ }\href {\doibase 10.1093/mnras/stt1025} {\bibfield  {journal} {\bibinfo  {journal} {Mon. Not. R. Astron. Soc.}\ }\textbf {\bibinfo {volume} {434}},\ \bibinfo {pages} {352} (\bibinfo {year} {2013}{\natexlab{a}})}\BibitemShut {NoStop}%
\bibitem [{\citenamefont {Chluba}(2013{\natexlab{b}})}]{chluba_distinguishing_2013}%
  \BibitemOpen
  \bibfield  {author} {\bibinfo {author} {\bibfnamefont {J.}~\bibnamefont {Chluba}},\ }\href {\doibase 10.1093/mnras/stt1733} {\bibfield  {journal} {\bibinfo  {journal} {Mon. Not. R. Astron. Soc.}\ }\textbf {\bibinfo {volume} {436}},\ \bibinfo {pages} {2232} (\bibinfo {year} {2013}{\natexlab{b}})}\BibitemShut {NoStop}%
\bibitem [{\citenamefont {Kite}\ \emph {et~al.}(2021)\citenamefont {Kite}, \citenamefont {Ravenni}, \citenamefont {Patil},\ and\ \citenamefont {Chluba}}]{kite_bridging_2021}%
  \BibitemOpen
  \bibfield  {author} {\bibinfo {author} {\bibfnamefont {T.}~\bibnamefont {Kite}}, \bibinfo {author} {\bibfnamefont {A.}~\bibnamefont {Ravenni}}, \bibinfo {author} {\bibfnamefont {S.~P.}\ \bibnamefont {Patil}}, \ and\ \bibinfo {author} {\bibfnamefont {J.}~\bibnamefont {Chluba}},\ }\href {\doibase 10.1093/mnras/stab1558} {\bibfield  {journal} {\bibinfo  {journal} {Mon. Not. R. Astron. Soc.}\ }\textbf {\bibinfo {volume} {505}},\ \bibinfo {pages} {4396} (\bibinfo {year} {2021})}\BibitemShut {NoStop}%
\bibitem [{\citenamefont {Cyr}\ \emph {et~al.}(2023)\citenamefont {Cyr}, \citenamefont {Kite}, \citenamefont {Chluba}, \citenamefont {Hill}, \citenamefont {Jeong}, \citenamefont {Acharya}, \citenamefont {Bolliet},\ and\ \citenamefont {Patil}}]{10.1093/mnras/stad3861}%
  \BibitemOpen
  \bibfield  {author} {\bibinfo {author} {\bibfnamefont {B.}~\bibnamefont {Cyr}}, \bibinfo {author} {\bibfnamefont {T.}~\bibnamefont {Kite}}, \bibinfo {author} {\bibfnamefont {J.}~\bibnamefont {Chluba}}, \bibinfo {author} {\bibfnamefont {J.~C.}\ \bibnamefont {Hill}}, \bibinfo {author} {\bibfnamefont {D.}~\bibnamefont {Jeong}}, \bibinfo {author} {\bibfnamefont {S.~K.}\ \bibnamefont {Acharya}}, \bibinfo {author} {\bibfnamefont {B.}~\bibnamefont {Bolliet}}, \ and\ \bibinfo {author} {\bibfnamefont {S.~P.}\ \bibnamefont {Patil}},\ }\href {\doibase 10.1093/mnras/stad3861} {\bibfield  {journal} {\bibinfo  {journal} {Mon. Not. R. Astron. Soc.}\ }\textbf {\bibinfo {volume} {528}},\ \bibinfo {pages} {883} (\bibinfo {year} {2023})}\BibitemShut {NoStop}%
\bibitem [{\citenamefont {Shibata}\ and\ \citenamefont {Sasaki}(1999)}]{shibata_black_1999}%
  \BibitemOpen
  \bibfield  {author} {\bibinfo {author} {\bibfnamefont {M.}~\bibnamefont {Shibata}}\ and\ \bibinfo {author} {\bibfnamefont {M.}~\bibnamefont {Sasaki}},\ }\href {\doibase 10.1103/PhysRevD.60.084002} {\bibfield  {journal} {\bibinfo  {journal} {Phys. Rev. D}\ }\textbf {\bibinfo {volume} {60}},\ \bibinfo {pages} {084002} (\bibinfo {year} {1999})}\BibitemShut {NoStop}%
\bibitem [{\citenamefont {Gow}\ \emph {et~al.}(2023)\citenamefont {Gow}, \citenamefont {Assadullahi}, \citenamefont {Jackson}, \citenamefont {Koyama}, \citenamefont {Vennin},\ and\ \citenamefont {Wands}}]{gow_non-perturbative_2023}%
  \BibitemOpen
  \bibfield  {author} {\bibinfo {author} {\bibfnamefont {A.~D.}\ \bibnamefont {Gow}}, \bibinfo {author} {\bibfnamefont {H.}~\bibnamefont {Assadullahi}}, \bibinfo {author} {\bibfnamefont {J.~H.~P.}\ \bibnamefont {Jackson}}, \bibinfo {author} {\bibfnamefont {K.}~\bibnamefont {Koyama}}, \bibinfo {author} {\bibfnamefont {V.}~\bibnamefont {Vennin}}, \ and\ \bibinfo {author} {\bibfnamefont {D.}~\bibnamefont {Wands}},\ }\href {\doibase 10.1209/0295-5075/acd417} {\bibfield  {journal} {\bibinfo  {journal} {EPL}\ }\textbf {\bibinfo {volume} {142}},\ \bibinfo {pages} {49001} (\bibinfo {year} {2023})}\BibitemShut {NoStop}%
\bibitem [{\citenamefont {Choptuik}(1993)}]{choptuik_universality_1993}%
  \BibitemOpen
  \bibfield  {author} {\bibinfo {author} {\bibfnamefont {M.~W.}\ \bibnamefont {Choptuik}},\ }\href {\doibase 10.1103/PhysRevLett.70.9} {\bibfield  {journal} {\bibinfo  {journal} {Phys. Rev. Lett.}\ }\textbf {\bibinfo {volume} {70}},\ \bibinfo {pages} {9} (\bibinfo {year} {1993})}\BibitemShut {NoStop}%
\bibitem [{\citenamefont {Evans}\ and\ \citenamefont {Coleman}(1994)}]{evans_critical_1994}%
  \BibitemOpen
  \bibfield  {author} {\bibinfo {author} {\bibfnamefont {C.~R.}\ \bibnamefont {Evans}}\ and\ \bibinfo {author} {\bibfnamefont {J.~S.}\ \bibnamefont {Coleman}},\ }\href {\doibase 10.1103/PhysRevLett.72.1782} {\bibfield  {journal} {\bibinfo  {journal} {Phys. Rev. Lett.}\ }\textbf {\bibinfo {volume} {72}},\ \bibinfo {pages} {1782} (\bibinfo {year} {1994})}\BibitemShut {NoStop}%
\bibitem [{\citenamefont {Niemeyer}\ and\ \citenamefont {Jedamzik}(1998)}]{niemeyer_near-critical_1998}%
  \BibitemOpen
  \bibfield  {author} {\bibinfo {author} {\bibfnamefont {J.~C.}\ \bibnamefont {Niemeyer}}\ and\ \bibinfo {author} {\bibfnamefont {K.}~\bibnamefont {Jedamzik}},\ }\href {\doibase 10.1103/PhysRevLett.80.5481} {\bibfield  {journal} {\bibinfo  {journal} {Phys. Rev. Lett.}\ }\textbf {\bibinfo {volume} {80}},\ \bibinfo {pages} {5481} (\bibinfo {year} {1998})}\BibitemShut {NoStop}%
\bibitem [{\citenamefont {Pi}\ and\ \citenamefont {Wang}(2023)}]{Pi_2023}%
  \BibitemOpen
  \bibfield  {author} {\bibinfo {author} {\bibfnamefont {S.}~\bibnamefont {Pi}}\ and\ \bibinfo {author} {\bibfnamefont {J.}~\bibnamefont {Wang}},\ }\href {\doibase 10.1088/1475-7516/2023/06/018} {\bibfield  {journal} {\bibinfo  {journal} {JCAP}\ }\textbf {\bibinfo {volume} {2023}},\ \bibinfo {pages} {018} (\bibinfo {year} {2023})}\BibitemShut {NoStop}%
\bibitem [{\citenamefont {Zhou}\ \emph {et~al.}(2020)\citenamefont {Zhou}, \citenamefont {Jiang}, \citenamefont {Cai}, \citenamefont {Sasaki},\ and\ \citenamefont {Pi}}]{PhysRevD.102.103527}%
  \BibitemOpen
  \bibfield  {author} {\bibinfo {author} {\bibfnamefont {Z.}~\bibnamefont {Zhou}}, \bibinfo {author} {\bibfnamefont {J.}~\bibnamefont {Jiang}}, \bibinfo {author} {\bibfnamefont {Y.-F.}\ \bibnamefont {Cai}}, \bibinfo {author} {\bibfnamefont {M.}~\bibnamefont {Sasaki}}, \ and\ \bibinfo {author} {\bibfnamefont {S.}~\bibnamefont {Pi}},\ }\href {\doibase 10.1103/PhysRevD.102.103527} {\bibfield  {journal} {\bibinfo  {journal} {Phys. Rev. D}\ }\textbf {\bibinfo {volume} {102}},\ \bibinfo {pages} {103527} (\bibinfo {year} {2020})}\BibitemShut {NoStop}%
\bibitem [{\citenamefont {Meng}\ \emph {et~al.}(2023)\citenamefont {Meng}, \citenamefont {Yuan},\ and\ \citenamefont {Huang}}]{meng_primordial_2023}%
  \BibitemOpen
  \bibfield  {author} {\bibinfo {author} {\bibfnamefont {D.-S.}\ \bibnamefont {Meng}}, \bibinfo {author} {\bibfnamefont {C.}~\bibnamefont {Yuan}}, \ and\ \bibinfo {author} {\bibfnamefont {Q.-G.}\ \bibnamefont {Huang}},\ }\href {\doibase 10.1007/s11433-022-2095-5} {\bibfield  {journal} {\bibinfo  {journal} {Sci. China Phys. Mech. Astron.}\ }\textbf {\bibinfo {volume} {66}},\ \bibinfo {pages} {280411} (\bibinfo {year} {2023})}\BibitemShut {NoStop}%
\bibitem [{\citenamefont {Byrnes}\ \emph {et~al.}(2024)\citenamefont {Byrnes}, \citenamefont {Lesgourgues},\ and\ \citenamefont {Sharma}}]{byrnes_robust_2024}%
  \BibitemOpen
  \bibfield  {author} {\bibinfo {author} {\bibfnamefont {C.~T.}\ \bibnamefont {Byrnes}}, \bibinfo {author} {\bibfnamefont {J.}~\bibnamefont {Lesgourgues}}, \ and\ \bibinfo {author} {\bibfnamefont {D.}~\bibnamefont {Sharma}},\ }\href@noop {} {\enquote {\bibinfo {title} {preprint},}\ } (\bibinfo {year} {2024}),\ \Eprint {http://arxiv.org/abs/2404.18475} {arXiv:2404.18475 [astro-ph.CO]} \BibitemShut {NoStop}%
\bibitem [{\citenamefont {{A. Kogut}}\ \emph {et~al.}(2011)\citenamefont {{A. Kogut}}, \citenamefont {{D.J. Fixsen}}, \citenamefont {{D.T. Chuss}}, \citenamefont {{J. Dotson}}, \citenamefont {{E. Dwek}}, \citenamefont {{M. Halpern}}, \citenamefont {{G.F. Hinshaw}}, \citenamefont {{S.M. Meyer}}, \citenamefont {{S.H. Moseley}}, \citenamefont {{M.D. Seiffert}}, \citenamefont {{D.N. Spergel}},\ and\ \citenamefont {{E.J. Wollack}}}]{a_kogut_primordial_2011}%
  \BibitemOpen
  \bibfield  {author} {\bibinfo {author} {\bibnamefont {{A. Kogut}}}, \bibinfo {author} {\bibnamefont {{D.J. Fixsen}}}, \bibinfo {author} {\bibnamefont {{D.T. Chuss}}}, \bibinfo {author} {\bibnamefont {{J. Dotson}}}, \bibinfo {author} {\bibnamefont {{E. Dwek}}}, \bibinfo {author} {\bibnamefont {{M. Halpern}}}, \bibinfo {author} {\bibnamefont {{G.F. Hinshaw}}}, \bibinfo {author} {\bibnamefont {{S.M. Meyer}}}, \bibinfo {author} {\bibnamefont {{S.H. Moseley}}}, \bibinfo {author} {\bibnamefont {{M.D. Seiffert}}}, \bibinfo {author} {\bibnamefont {{D.N. Spergel}}}, \ and\ \bibinfo {author} {\bibnamefont {{E.J. Wollack}}},\ }\href {\doibase 10.1088/1475-7516/2011/07/025} {\bibfield  {journal} {\bibinfo  {journal} {JCAP}\ }\textbf {\bibinfo {volume} {2011}},\ \bibinfo {pages} {025} (\bibinfo {year} {2011})}\BibitemShut {NoStop}%
\bibitem [{\citenamefont {Dent}\ \emph {et~al.}(2012)\citenamefont {Dent}, \citenamefont {Easson},\ and\ \citenamefont {Tashiro}}]{dent_cosmological_2012}%
  \BibitemOpen
  \bibfield  {author} {\bibinfo {author} {\bibfnamefont {J.~B.}\ \bibnamefont {Dent}}, \bibinfo {author} {\bibfnamefont {D.~A.}\ \bibnamefont {Easson}}, \ and\ \bibinfo {author} {\bibfnamefont {H.}~\bibnamefont {Tashiro}},\ }\href {\doibase 10.1103/PhysRevD.86.023514} {\bibfield  {journal} {\bibinfo  {journal} {Phys. Rev. D}\ }\textbf {\bibinfo {volume} {86}},\ \bibinfo {pages} {023514} (\bibinfo {year} {2012})}\BibitemShut {NoStop}%
\bibitem [{\citenamefont {Yoo}\ \emph {et~al.}(2021)\citenamefont {Yoo}, \citenamefont {Harada}, \citenamefont {Hirano},\ and\ \citenamefont {Kohri}}]{Yoo:2020dkz}%
  \BibitemOpen
  \bibfield  {author} {\bibinfo {author} {\bibfnamefont {C.-M.}\ \bibnamefont {Yoo}}, \bibinfo {author} {\bibfnamefont {T.}~\bibnamefont {Harada}}, \bibinfo {author} {\bibfnamefont {S.}~\bibnamefont {Hirano}}, \ and\ \bibinfo {author} {\bibfnamefont {K.}~\bibnamefont {Kohri}},\ }\href {\doibase 10.1093/ptep/ptaa155} {\bibfield  {journal} {\bibinfo  {journal} {PTEP}\ }\textbf {\bibinfo {volume} {2021}},\ \bibinfo {pages} {013E02} (\bibinfo {year} {2021})}\BibitemShut {NoStop}%
\end{thebibliography}%

\appendix

\section{Calculation of PBH abundance based on compaction function}\label{PBHs_cal_details}

In this appendix, we provide details on calculating the abundance of PBHs corresponding to a given primordial power spectrum using the compaction function and consider the effects of profile and threshold \cite{young_calculating_2014,musco_threshold_2019,Escrivà_2021,Yoo:2020dkz,gow_non-perturbative_2023}.

The compaction function is considered for calculating the abundance of PBHs, but the threshold of it to form PBHs depends on the profile. Numerical simulations and analysis show that the volume-averaged compaction function $\bar{\mathcal{C}}(w)$ can also be used as a criterion for forming PBHs and provide a profile-independent threshold. The averaged critical compaction function threshold can be obtained from the following expression:
\begin{equation}
\bar{\mathcal{C}}_\mathrm{th}(w)=a+b ~ \mathrm{arctan}(cw^d),
\end{equation}
where the parameters are $a= -0.140381$, $b = 0.79538$, $c = 1.23593$, $d = 0.357491$ and equation of state $w =1/3$ in radiation domination.

To relate this result to the compaction function, we consider in the overdense region with radius $r$ in perturbative perspective, the typical profile of a given Gaussian random field $\zeta_G(r)$ with a high peak is
\begin{equation}
    \begin{split}
    \hat{\zeta}_G(r)
    &=\Tilde{\mu}_2\Bigg[\frac{1}{1-\gamma_3^2}\Big(\psi_1(r)+\frac{1}{3}R_3^2\Delta\psi_1(r)\Big)\\
    &\hspace*{3em}-\tilde{k}_3^2\frac{1}{\gamma_3(1-\gamma_3^2)}\Big(\gamma_3^2\psi_1(r)+\frac{1}{3}R_3^2\Delta\psi_1(r)\Big)\Bigg],
    \end{split}
\end{equation}
where $\tilde{\mu}_2$ and $ \tilde{ k}_3$ represent the peak's height and width respectively. The other statistic parameters can be obtained from the perturbation's power spectrum, and `G' denotes Gaussian perturbation. Thus, we can obtain the threshold of the compaction function $\mathcal{C}_{\rm{th}}$ from the threshold of the averaged compaction function $\bar{\mathcal{C}}_{\rm{th}}$ and the typical profile $\hat{\zeta}(r)=F(\hat{\zeta}_G)$. In this paper, we use the simplified typical profile $\hat{\zeta}_G=\Tilde{\mu}_2\psi_1(r)$ when taking the average of $ \tilde{ k}_3$,
where
\begin{equation}
    \psi_1=\frac{\int\frac{{\rm{d}} k}{k}k^2\frac{{\rm{sin}} (kr)}{kr}\mathcal{P}_{\zeta}(k)}{\int\frac{{\rm{d}} k}{k}k^2\mathcal{P}_{\zeta}(k)},
\end{equation} 
and $\Tilde{\mu}_2$ can be considered as a 
Gaussian random variable. Further, we consider the compaction function can be expanded by linear compaction functions $\mathcal{C}_\ell $ as:
\begin{equation}
\mathcal{C}(r)=\mathcal{C}_\ell-\frac{3}{8} \mathcal{C}_\ell^2,
\end{equation}
and $\mathcal{C}_\ell$ can be written in terms of $\zeta$ as $\mathcal{C}_\ell=-\frac{4}{3}r\zeta'$ with the derivative $\prime$ with respect to $r$. The threshold of the linear compaction function is 
\begin{equation}
C_{\ell,\rm{th}}=\frac{4}{3}\left(1-\sqrt{1-\frac{3}{2} C_{\rm{th}}}\right).
\end{equation}

Because the derivative $\zeta^\prime (r)$ depends on $r$, we expect to rewrite it in terms of $r$ and $\zeta_G$ in a local transformation, thus the linear compaction function can be expressed as:
\begin{equation}
    \mathcal{C}_\ell = -\frac{4}{3} r \zeta'_G \mathcal{J}(\zeta_G),
\end{equation}
where $\mathcal{J}(\zeta_G)= {\mathrm{d} \zeta}/{\mathrm{d} \zeta_G}$, in Gaussian case, $\mathcal{J}(\zeta_G)=1$. To express the probability distribution function of $\mathcal{C}_\ell$, we assume two Gaussian random variable $ X = r\zeta_{G}^{\prime} $, $ Y = \zeta_{G} $ they follow the joint probability distribution:
\begin{equation}
    \mathbb{P}(X,Y)=\frac{1}{2\pi\sqrt{\det(\boldsymbol{\Sigma})}}\exp(-\frac{\boldsymbol{V}^T\boldsymbol{\Sigma}^{-1}\boldsymbol{V}}{2}),
\end{equation}
where $\boldsymbol{V}^T=(X,Y),\boldsymbol{\Sigma}=\begin{pmatrix}
    \Sigma_{XX} &\Sigma_{XY} \\
    \Sigma_{XY}&\Sigma_{YY}
\end{pmatrix}$.
The covariance matrix $\boldsymbol{\Sigma}$ can be calculated from the power spectrum of Gaussian perturbation $\mathcal{P}_{\zeta_G}(k)$:
\begin{align}
    \Sigma_{XX} &= \int  {\rm{d}} (\operatorname{ln}k)(kr)^2 [\frac{{\rm{d}}j_0}{{\rm{d}}z}(kr)]^2T^2 ( k, r) \mathcal{P}_{\zeta_G}(k), \\
    \Sigma_{YY} &= \int {{\rm{d}}} (\operatorname{ln}k)j^2_0(kr)T^2 ( k, r)\mathcal{P}_{\zeta_G}(k), \\
    \Sigma_{YX} &= \Sigma_{XY}=\int {\rm{d}} (\operatorname{ln}k) (kr)j_0(kr)\frac{{\rm{d}}j_0}{{\rm{d}}z}(kr)T^2 ( k, r)\mathcal{P}_{\zeta_G}(k),
\end{align}
where $j_0=(\sin x)/x$ is the spherical Bessel function, and $T ( k, \tau)$ is the transfer function in radiation domination:
\begin{equation}
	T ( k, \tau) = 3 \frac{\operatorname{s i n} \left( k \tau/ \sqrt{3} \right) - ( k \tau/ \sqrt{3} ) \operatorname{c o s} \left( k \tau/ \sqrt{3} \right)} {( k \tau/ \sqrt{3} )^{3}}.
\end{equation}
In the equation, we take $r=\tau=r_m$, where $r_m$ is the location of the first maximum value of $\mathcal{C}(r)$.

Therefore, the linear compaction function follows the distribution:
\begin{equation}
        \mathbb{P}(\mathcal{C}_\ell)  =\int {\rm{d}} X\int {\rm{d}} Y \mathbb{P} (X,Y)\delta_\mathrm{D}[\mathcal{C}_\ell-\mathcal{C}_\ell(X,Y)].
\end{equation}
With the help of the 2D PDF, the possibility of PBH formation from perturbation is determined.

The PBH mass $M$ is related to the linear compaction $\mathcal{C}_\ell$ by the critical collapse
\begin{equation}
    M=\mathcal{K}M_H(\mathcal{C}_\ell-\frac{3}{8}\mathcal{C}_\ell^2-\mathcal{C}_{\rm{th}})^{\gamma}.
\end{equation}
We take $\gamma=0.357$ and $\mathcal{K}=4$
and total abundance of PBH is:
\begin{equation}
    \begin{split}
    \beta_\mathrm{total}
    &=\int^{M_{\rm{max}}}_0 {\rm{d}}(\operatorname{ln}M) \beta(M) \\
    & =\int^{M_{\rm{max}}}_0 {\rm{d}}(\operatorname{ln}M) \frac{\mathcal{K}(\mathcal{C}_\ell-\frac{3}{8}\mathcal{C}_\ell^2-\mathcal{C}_\mathrm{th})^{\gamma+1}}{\gamma
    (1-\frac{3}{4}\mathcal{C}_\ell)}\mathbb{P}(\mathcal{C}_\ell),
    \end{split}
\end{equation}
where $M_{\mathrm{max}}$ is the mass corresponding to the maximum value of $\mathcal{C}_\ell$.

\end{document}